\documentclass[%
 reprint,
superscriptaddress,
 amsmath,amssymb,
 aps,
]{revtex4-2}

\usepackage{comment}
\usepackage{graphicx}% Include figure files
\usepackage{booktabs}
\usepackage{dcolumn}% Align table columns on decimal point
\usepackage{bm}
\usepackage[dvipsnames]{xcolor}
\usepackage{newtxmath}
\usepackage{mathtools}
\usepackage{physics}
\usepackage{float}
\usepackage[caption = false]{subfig}
\usepackage{tabularx}
\usepackage{pifont}

\newcommand{\be}{\begin{equation}}
\newcommand{\ee}{\end{equation}}

\def\l{{\Lambda}}
\def\lf{{\Lambda_f}}
\def\K{{\mathcal{K}}}
\def\Q{{\mathcal{Q}}}

\begin{document}

\preprint{APS/123-QED}

\title{Quantum criticality and non-Fermi liquids: \\ 
the Wilsonian renormalization group perspective}
%3) 

\author{Mateusz Homenda  }
\email{mateusz.homenda@fuw.edu.pl}
\author{Pawel Jakubczyk } 
\email{pawel.jakubczyk@fuw.edu.pl}
\affiliation{Institute  of Theoretical Physics, Faculty of Physics, University of Warsaw, Pasteura 5, 02-093 Warsaw, Poland} 
\author{Hiroyuki Yamase } 
\email{yamase.hiroyuki@nims.go.jp}
\affiliation{Research Center for Materials Nanoarchitectonics, National Institute for Materials Science, Tsukuba 305-0047, Japan}

\date{\today}

\begin{abstract}
We develop a theoretical framework based on the nonperturvative renormalization group (RG) in the one-particle irreducible (Wetterich)  formulation to tackle the interplay of coupled fermionic and order-parameter fluctuations at metallic quantum critical points (QCPs) with ordering wavevector $\vec{Q}=\vec{0}$. We consistently treat the dynamical emergence of the Landau damping of the bosonic mode and non-Fermi liquid scaling of fermions upon lowering the cutoff scale. The loop integrals of the present theory involve only contributions from fluctuations above the cutoff scale, which protects the system from developing singular bosonic interactions. 
We emphasize the importance of the nontrivial relative scaling of the bosonic and fermionic cutoffs $\Lambda$ and $\Lambda_f$, which we fix by analyzing the RG flow of the scale-dependent ordering wave-vector $\vec{Q}_\Lambda$.
Upon neglecting Fermi self-energy in the loop integrals of the functional RG flow, we identify a non-Fermi liquid RG fixed point and recover the features obtained  earlier within RPA-type approaches.  
In a subsequent step,  
we self-consistently include the scaling of the self-energy and the Yukawa coupling. We find a generic instability of the non-Fermi liquid RG fixed point. This implies, at least at this truncation level,  absence of the QCP with $\vec{Q}=\vec{0}$ and development of a first-order phase transition or a phase characterized by $\vec{Q}\neq\vec{0}$.

%In particular we demonstrate how results characteristic for the RPA-type fixed-point scaling are recovered by a questionable procedure of removing the fermionic cutoff much faster than the bosonic one. Departure from this paradigm in a simple truncation neglecting the self-energy in loop integrals  drives the system to a non-Fermi liquid RG fixed point characterized by scaling properties different from those anticipated within the standard approaches. We find however that minimal extensions including the renormalization of self-energy and interaction coupling tend to push the flowing ordering wavevector to $\vec{Q}\neq \vec{0}$, which can be avoided only by imposing severe additional constraints on the fixed point. 

%enforcing a particular relation between $\Lambda_b$ and $\Lambda_f$   

%In particular the scaling exponent $\alpha$ for the Fermi self-energy acquires a value lower than the anticipated $\alpha\approx 0.66$, while the bosonic dynamical exponent $z\in [2,3)$. We demonstrate that inclusion of self-energy in the loops of the RG flow introduces strong constraints on the RG flow \dots  

\end{abstract}

\maketitle

\section{\label{sec:INTRO} Introduction} 
Quantum criticality of metallic systems and the associated breakdown of Landau's Fermi liquid theory have been receiving enormous interest over the last decades due to their relevance to high-$T_c$ cuprate superconductors \cite{Keimer_2015} and other strongly-correlated materials. 

The earliest  theoretical investigations of these phenomena \cite{Moriya_book, Hertz_1976, Millis_1993} (referred to as Hertz-Millis theory) departed from the paradigms of Wilsonian momentum-shell  renormalization group applied to an effective bosonic action. The latter was obtained by completely integrating out the fermionic degrees of freedom out of the theory, giving rise to a bosonic self-energy involving the so-called Landau-damping term, which describes damping of the propagating order-parameter mode due to scattering off the particle-hole excitations. Later  studies revealed inadequacies of the Hertz-Millis approach in two-dimensional systems. The reason for these  is understood by noting that, in addition to the order-parameter mode, the system hosts electronic excitations across the Fermi level, which are also soft. Integrating out some of the massless modes and leaving the others in the effective theory is unjustified from the point of view of Wilsonian RG theory. Several studies spread over years \cite{Chubukov_2004, Chubukov_2004_2, Belitz_2005,  Lohneysen_2007, Thier_2011, Metlitski_2010, Sachdev_2011} note that the procedure yields singular effective bosonic interactions, which cannot be handled within the Hertz-Millis framework. 

Theoretical developments which followed, led over years to the consensus that a satisfactory effective theory of metallic quantum criticality should explicitly keep the low-energy fermionic degrees of freedom. Several seminal papers  pursued  this path \cite{Lee_2009, Maslow_2010, Metlitski_2010, Mross_2010, Dalidovich_2013,
Torroba_2014, Mandal_2015, Holder_2015Jul,  Punk_2016, Lee_2018, Damia_2019, Damia_2020, Saterskog_2021, Zhang_2023, Mayrhofer_2024, Kukreja_2024}, implementing the tools of field-theoretic RG, a variety of different regularizations, and taking advantage of affinity to an earlier studied problem of fermions coupled to a $U(1)$ gauge field.  A common feature of these approaches is that the bosonic propagator becomes dressed in the singular Landau-damping term before performing any loop integrals of the coupled Bose-Fermi theory. The approach yields a picture with some surprising features: the standard $1/N$ expansion ($N$ being the number of Fermi flavors) is questionable \cite{Lee_2009} since diagrams of any loop order contribute also in the limit $N\to \infty$; the one-loop result for the dynamical exponent of fermions acquires a tiny correction only at three-loop level, while the corresponding result for the bosonic dynamical exponent $z_b$ is robust at two- and three-loop order. However, an unexpected thing happens at four-loop \cite{Holder_2015Jul, Holder_2015Dec, Ye_2022}, where a   nonrenormalizable divergence is found, which, as far as we are aware, was not clearly interpreted up to now.    

Studies of coupled Fermi-Bose theories of metallic quantum criticality from the Wilsonian RG perspective are more scarce \cite{Drukier_2012, Fitzpatrick_2013,  Ridgway_2015, Trott_2018, Sheerin_2024}. A key question in this approach concerns the way Landau damping should emerge in course of the RG flow. As emphasized in Ref.~\cite{Fitzpatrick_2013}, its standard form cannot be continuously generated at finite scales upon reducing the cutoff around the Fermi surface, since (as visible even from evaluating the bare Fermi loop), the Landau damping term arises in a singular way exclusively from fermions directly at the Fermi level. From the Wilsonian RG point of view, therefore, it is illogical to use the standard Landau-damped Bose propagator in RG theory of coupled Bose-Fermi systems because it appears only after the fermions become completely removed from the theory. On the other hand,  observable physical quantities %such as, for example, the dynamical exponent of the Bose field 
depend on the precise form of the Bose propagator. A notable case that permits an unbiased Wilsonian RG analysis without addressing the Landau damping is the chiral non-Fermi liquid~\cite{Sur_2014}, where  exact critical exponents can be deduced.  

An approach to this generic problem pertinent to instabilities at ordering wavevector $\vec{Q}=\vec{0}$ was recently proposed by us \cite{Homenda_2024} implementing the modern formulation of Wilsonian nonperturbative RG relying on the Wetterich equation \cite{Wetterich_1993, Morris1994}, applied to a system of coupled fermionic and order-parameter fluctuations. This leads to a generalization of the Hertz action, which recovers the standard form only in an undesirable procedure of integrating out fermions before the bosons. On the other hand, it encompasses the same physics and yields a consistent picture, free of singular effective interactions, where fermions and bosons are treated on equal footing and integrated out in parallel. 
Ref.~\cite{Homenda_2024} concentrated on bosonic properties and the emergent order-parameter action neglecting the Fermi self-energy terms. 
The present paper constitutes a continuation of that avenue, accounting, {\it inter alia}, for those effects. We additionally elaborate on the relation between the two cutoff scales present in our approach and clarify how, at RPA-type level, it leads to the dynamical exponent $z_b=3$. This picture however breaks down, when both the self-energy effects and the flow of the Yukawa vertex are taken into account.   

Throughout the present paper we restrict to the case of instabilities occurring at zero ordering wavevector $\vec{Q}$. 
The primary one-loop prediction concerning the bosonic properties in this case is the mean-field nature of the critical exponents at $T=0$. This can be understood \cite{Hertz_1976, Millis_1993, Nagaosa_1998, Sachdev_2011} by considering that integrating fermions out generates a bosonic action in effective dimensionality $d_{\rm eff} \coloneqq d + z_b$, where $d$ is the spatial dimension and $z_b$ the dynamical critical exponent that  characterizes scaling of  correlation time with correlation length, $\tau \sim \xi^{z_b}$.  According to the one-loop result, the low-energy order-parameter dynamics is governed by the Landau damping term $\sim|q_0|/|\vec{q}|$ in the propagator, where $q_0$ and $\vec{q}$ denote bosonic Matsubara frequency and momentum, respectively.  By comparing the Landau damping term with the standard quadratic term $|\vec{q}|^2$ in the boson propagator, one obtains $z_b = 3$. In consequence, for two-dimensional systems  $d_{\rm eff} = 5$, which exceeds the upper critical dimension $d_{\rm up} = 4$. These are the main one-loop characteristics of the bosonic sector at $T=0$. In reference to fermions, we will use the phrases "one-loop" and "RPA-type" interchangeably. Concerning the fermionic properties, % at quantum critical points with $\vec{Q}=\vec{0}$, 
we invoke the broadly known one-loop result \cite{Sachdev_2011} stating that the Fermi propagator $G_f$ for small frequencies $k_0$ behaves as:
\begin{equation}
\label{alpha_fer}
    G_f^{-1}(k_0, |\vec{k}| = k_f) \sim -i \; {\rm sgn}(k_0) |k_0|^{\alpha} \; ,
\end{equation}
where $\vec{k}$ denotes the fermionic wavevector and 
$k_f$ is the Fermi momentum. 
At one loop level one obtains $\alpha = 2/3$, which implies non-Fermi liquid behavior. 

We emphasize that the precise values of $z_b$ and $\alpha$ in $d=2$ have never been  well established  in literature beyond the robust one-loop result. The situation is  intriguing considering that perturbation theory corrects the value $\alpha = 2/3$ only at three-loop level \cite{Metlitski_2010}, while, as already mentioned, a four-loop calculation leads to contributions to $z_b$ that have no clear interpretation \cite{Holder_2015Jul} and rather point towards foundational problems of the applied approach.
The picture provided by the quantum Monte Carlo simulations is also far from definitive with regard to numbers. The study of peculiar electronic-nematic systems shows $z_b \approx 2$ scaling \cite{Schattner_2016} while simulations of Ref.~\cite{Xu_2017}  yield a bosonic susceptibility  inconsistent with the Hertz form $\sim |q_0|/|\vec{q}|$. These deviations were however later attributed to finite size \cite{Klein_2017} and thermal effects \cite{Klein_2020, Xu_2020}. Interestingly, simulations of the $XY$ model of itinerant ferromagnetism provide  evidence for $z_b = 2$ and $\alpha = 1/2$ \cite{Liu_2022}. The latter result is also consistent with a recent renormalization group analysis of isotropic Fermi gas interacting with U(1) gauge field, which is expected to belong to the same universality class \cite{Sheerin_2024}. Notably, at Hertz-Millis level,  these values of $z_b$ and $\alpha$ are characteristic  to  instabilities featuring ordering wavevector  $\vec{Q}\neq \vec{0}$. For more recent insights into this problem see \cite{Schlief_2017}.

In the present work we develop an  approach based on a  starting point, in which the propagators are only gradually dressed upon decreasing the RG cutoff scale and the problems encountered in the previous studies are absent.     

The outline of the  paper is the following: In Sec.~II we introduce the model and discuss the applied functional RG approach. In Sec.~III we analyze non-self-consistent truncations of the Wetterich equation leading to a generalization of the Hertz action, flow of the self-energy to a  non-Fermi liquid state, and the associated RG fixed point. We also demonstrate how the standard Hertz action is recovered by a dubious procedure of scaling the cutoff on fermionic degrees of freedom to zero much faster than the one applied on the bosonic ones. In Sec.~IV we self-consistently include the flow of the self-energy and the Yukawa vertex. We demonstrate non-existence of the non-Fermi liquid  RG fixed point at the considered truncation level. In Sec.~IV.C we  speculate on the possibility of reemergence of the second-order quantum phase transition with $\vec{Q}=\vec{0}$ if additionally four-point functions involving both fermions and bosons were taken into account. We summarize the paper in Sec.~V, and technical details are presented in the Appendices. 

\section{\label{sec:MODEL}Model and functional renormalization group}
We analyze the coupled flows of fermionic and order-parameter fluctuations employing a modern version of the Wilsonian RG based on the Wetterich equation \cite{Wetterich_1993, Morris1994}. The latter is an exact functional flow equation for the one-particle-irreducible vertex functions equipped with a (momentum)  cutoff at scale $\Lambda$. This nonperturbative framework has, in recent years, allowed for groundbreaking progress in addressing several problems hardly accessible to traditional approaches based on perturbation theory, and sometimes completely beyond their reach. Notable  achievements of this method within the field of condensed matter theory include identifying the strong-coupling fixed points for the Kardar-Parisi-Zhang problem in $d\geq 1$ \cite{Canet_2010, Fontaine_2023}, resolution of the long-standing problem of dimensional reduction and its breaking in disordered systems  \cite{Tissier_2011, Tarjus_2025}, finding entirely new multicritical RG fixed points for the $O(N)$ models in $d=3$ \cite{Yabunaka_2017}, and invalidation \cite{Chlebicki_2021} of the predictions of perturbative approaches concerning non-analyticity of the critical exponents as function of $d$ and $N$. We refer to \cite{Kopietz_book, Dupuis_2021} for reviews. In view of the present problem, the important characteristic of the functional Wetterich approach is that it allows for efficient analysis of coupled flows of interacting degrees of freedom without specifying any preimposed  form of the flowing quantities and consistently arranging {\it all} of the fluctuations contributing to the partition function according to the decreasing energy scale.

\subsection{The bare action and the flowing action}
As our starting point we consider a two-dimensional system of fermions, described by Grassmann fields $\{\psi, \bar{\psi}\}$, coupled to the real scalar  bosonic field $\phi$ by a Yukawa-type interaction. The bare action $S[\psi, \bar{\psi}, \phi]$ is defined at a microscopic scale $\Lambda_{UV}$ and involves three terms \cite{Sachdev_2011}:
\begin{eqnarray}
    \begin{aligned}
&S[\psi, \bar{\psi}, \phi] = S_b + S_f + S_{fb} \\
    &S_f = \sum_{\sigma} \int_{K} \bar{\psi}_{K, \sigma} \left(- i k_0 + \xi_{\vec{k}} \right) \psi_{K, \sigma}\\
        &S_b = \int_Q (m_b^2 + |\vec{q}|^2 + q_0^2)\phi_{Q}\phi_{-Q}  + u \int_{x, t} (\phi(x, t))^4 \\
        &S_{fb} = g  \sum_{\sigma} \int_{Q, K} \phi_Q \bar{\psi}_{K , \sigma} \psi_{K-Q , \sigma}   \; , 
    \end{aligned} 
 \label{bare_action}   
\end{eqnarray}
where $m_b^2\,, u$, and $g >0$ are the bare parameters of the system. Here $K = \{ k_0, \vec{k} \}$ encompasses fermionic frequency and momentum, while $Q = \{ q_0, \vec{q} \}$ stands for their bosonic counterparts. We also define an abbreviation for integrals, $\int_K \coloneqq (\tfrac{1}{2 \pi})^3 \int {\rm d} k_0 \int {\rm d}^2 k $. Finally we consider the spherical Fermi surface $\xi_{\vec{k}} = \tfrac{1}{2 m_f} (|\vec{k}|^2 - k_f^2)$.

The bosonic part $S_b$ taken alone describes the phase transition of  the 2$d$ quantum Ising type, which is governed by the RG fixed point of the classical 3$d$ Ising universality class. In the Hertz-Millis framework (emerging after integrating out the fermionic fields $\{\psi, \bar{\psi}\}$) this situation is different due to % coupling to fermions and 
appearance of the Landau damping term $\sim |q_0|/|\vec{q}|$, which effectively  describes interactions of the bosonic mode with  particle-hole excitations. The $T=0$ critical behavior of such a damped boson field is controlled by the Gaussian fixed point. 

\begin{figure}
    \centering
    \includegraphics[width=1.\linewidth]{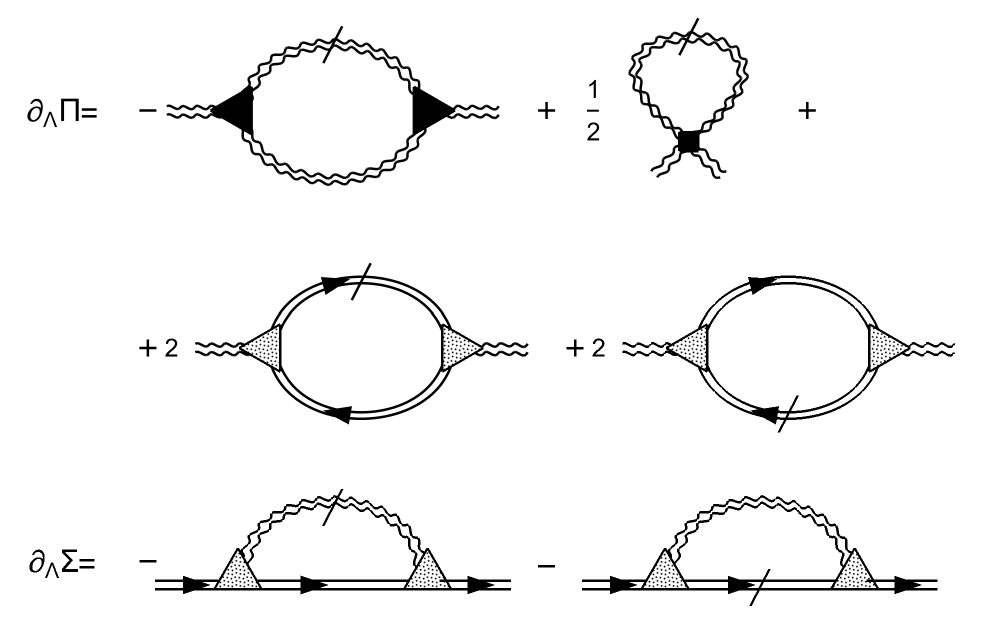}
    \caption{Feynman diagrams representing the flow of the self-energies. The wiggly (straight) line denotes the boson (fermion) propagator $G_{b/f}$ dressed by fluctuations down to the cutoff scale $\Lambda$ (or $\Lambda_f$). The strikethrogh indicates the single scale propagator $\mathcal{S}_{b/f}$. Contributions involving 4-point functions with fermionic legs were omitted.}
    \label{fig:diagrams}
\end{figure}

In order to consistently formulate the theory and in particular describe the gradual generation of Landau damping in course of the flow, we add the scale-dependent cutoff functions $R_b(Q,\Lambda)$ and $R_f(K,\Lambda_f)$ to the bosonic and fermionic inverse propagators, which deforms the quadratic part of the action according to:
\begin{eqnarray}
&S_{f,\lf} = S_f + \sum_{\sigma} \int_K  R_f \;\bar{\psi}_{K, \sigma} \psi_{K, \sigma} \\
&S_{b,\l} = S_b + \int_Q  R_b \;\phi_{Q} \phi_{-Q}\;.
\end{eqnarray}
Here $\l$ is the IR cut-off scale for bosonic momenta and $\lf(\l)$ describes the fermionic cutoff scale, i.e., the distance to the Fermi surface up to which the fermionic dispersion is modified. The relation between $\lf$ and $\l$ will be discussed later. The introduction of cut-off functions regularizes all infrared divergencies and at the same time allows for the derivation of the Wetterich equation, which governs the flow of the effective action $\Gamma_{\l}[\psi, \bar{\psi},\phi]$ upon varying the cutoff scale --- see the Appendix A. The  functional $\Gamma_{\l}[\psi, \bar{\psi},\phi]$ interpolates  between the bare action $S[\psi, \bar{\psi},\phi]$, recovered in the limit $\Lambda\to\Lambda_{UV}$, and the full effective action (i.e. the Gibbs free energy)  for $\l \to 0$. Taking consecutive derivatives of the Wetterich equation with respect to fields yields exact  flow equations for the (1-particle irreducible)  correlation functions 
\begin{eqnarray}
\Gamma^{(n)}_{\alpha_1 ... \alpha_n} &\coloneqq& \dfrac{\delta^{n} \Gamma_{\l}[\psi, \bar{\psi},\phi]}{ \delta \alpha_{1} ... \delta \alpha_{n}} \,  ,\;\; \alpha_i \in \{ \phi_{Q_i}, \psi_{K_i , \sigma_i}, \bar{\psi}_{K_i , \sigma_i} \}\,. 
\end{eqnarray} 
In this work, we focus on the flow of fermionic and bosonic self-energies, $\Sigma(K)$ and $\Pi(Q)$, directly related to $\Gamma^{(2)}_{\alpha_1,\alpha_2}$; the corresponding flow equations are given in the Appendix A --- see Eq.~(\ref{app1:Gamma2bos}) and Eq.~(\ref{app1:Gamma2fer}), and are represented by the Feynman diagrams depicted in Fig.~\ref{fig:diagrams}. For future reference we also introduce the regularized, scale-dependent propagators 
\begin{eqnarray}
G_b(Q) &\coloneqq& \left[\Gamma^{(2)}_{\phi\phi} (Q)+R_b(Q)\right]^{-1}  \\ 
&=&\frac{1}{m_b^2 + |\vec{q}|^2 + q_0^2 + \Pi(Q) + R_b(Q)} \;,\nonumber \\
        G_f(K) &\coloneqq& \left[\Gamma^{(2)}_{\psi\bar{\psi}}(K,\sigma)+R_f(K,\sigma)\right]^{-1} \\ 
        &=& \frac{1}{- i k_0 + \xi_{\vec{k}} + \Sigma(K) + R_f(K)}\;,  \nonumber
\end{eqnarray} 
as well as the single-scale propagators
\begin{eqnarray}
        \mathcal{S}_i &\coloneqq& - (\partial_{\l_i}R_i)
        \,(G_i)^2 \; , \qquad i \in \{b, f\} \; .
\end{eqnarray}  
Above we suppressed the scale-dependencies for clarity.
Our notation is fully consistent with the one of Ref. \cite{Homenda_2024} (see also \cite{Obert_2013, Obert_2014_PhD}).
For a thorough discussion of the Wetterich framework we refer to \cite{Berges_2002, Delamotte_2012, Kopietz_book, Dupuis_2021}.

\section{Generalized Hertz action \\and the non-Fermi liquid fixed point}

In perturbation theory the standard Landau damping is generated at one-loop level, and does not require accounting for self-energy contributions. In what follows we explain how that result is recovered in our present  framework by a questionable procedure of integrating fermionic degrees of freedom via the flow much faster than the bosonic ones and how improvement can be achieved in the present methodology. We subsequently present a detailed discussion concerning the structure of the regularized bubble diagram (second line in Fig. \ref{fig:diagrams}) and the relation between the scales $\Lambda$ and $\Lambda_f$, which, in our approach, is fixed by the magnitude of the flowing ordering wave-vector $\vec{Q}_\Lambda$. This characteristic is crucial for values of scaling exponents.

Having in mind the goal of recovering the standard one-loop theory from the Wetterich flow and setting up a starting point for more elaborate analysis, we propose a  truncation of the Wetterich equation, which is summarized as follows: \\
1. We neglect the Fermi self-energy in the loop integrals of the flow equations. \\
%2. We project the Fermi self-energy on the Fermi level, i.e. $\Sigma(k_0, \vec{k})\longrightarrow \Sigma(k_0, k_f)$.\\ 
2. We disregard the flow of the Yukawa coupling and the higher order interactions generated by the RG flow except for the bosonic self-interaction $u$, which is treated as a flowing local coupling. 

Truncations reaching beyond this level will be discussed in Sec.~IV.
Note that there is no preimposed parametrization of the momentum and frequency dependencies of the flowing propagators. As we discuss below, the truncation allows for making contact with the Hertz-Millis framework at any time by scaling $\Lambda_f$ to zero before $\Lambda$. Presence of the fermionic cutoff prevents generation of singular bosonic interactions. With the above specified truncation, we obtain the following flow equations for the self-energies: 

\begin{eqnarray} \label{eq:pi}
\begin{aligned}
    \partial_{\l}\Pi(Q') =  &\!\begin{multlined}[t]
        - \int_Q \mathcal{S}_b(Q) 
        \Gamma^{(3)}_{\phi \phi \phi} 
        G_{b}(Q+ Q') 
        \Gamma^{(3)}_{\phi \phi \phi} \\
    + \frac{1}{2} \int_Q \mathcal{S}_b(Q) \Gamma^{(4)}_{\phi \phi \phi \phi}
    \end{multlined} \\
    &+ 2 g^2 (\partial_\Lambda \Lambda_f) \int_K \mathcal{S}_{f,0}(K) 
    G_{f,0}(K + Q')   \\
    &+ 2 g^2 (\partial_\Lambda \Lambda_f) \int_K \mathcal{S}_{f,0}(K) 
    G_{f,0}(K - Q') \; ,
\end{aligned}
\end{eqnarray}
\begin{eqnarray} \label{eq:sigma}
\begin{aligned}
    \partial_{\l}\Sigma(K') =  
    &- g^2 \int_Q \mathcal{S}_b(Q) 
    G_{f,0}(K' - Q) \\
    &- g^2 (\partial_\Lambda \Lambda_f)  \int_Q \mathcal{S}_{f,0}(K' - Q) 
    G_b(Q) \;,
\end{aligned}
\end{eqnarray} 
where 
\begin{align} 
G_{f, 0}(K) \coloneqq  \frac{1}{- i k_0 + \xi_{\vec{k}} + R_f(K)}\,,\;\; \mathcal{S}_{f,0}:= -(\partial_{\lf}R_f)
        \,(G_{f,0})^2 \,.
\end{align}
%The diagrammatic interpretation of these equations is given at the top of Fig.\dots. From now on we omit the spin index in fermionic self-energies and propagators.  
Throughout the paper we concentrate on the quantum critical state, which is achieved by tuning the bare boson mass such that it scales to zero in the limit $\Lambda\to 0$. The flow equation for the mass $m_b^2=\Pi(Q=0)$ includes contributions from the fermion 
loop, boson bubble, and boson tadpole. In fact, we checked that disregarding the purely bosonic contributions does not affect our results. 

The present framework is  different from the approaches relying on perturbation theory. In the latter, the one-loop boson self-energy is computed using the bare fermion propagator, leading to the $|q_0|/|\vec{q}|$ Landau damping term. This is subsequently  
inserted into the one-loop fermion self-energy diagram, which yields the $|k_0|^{2/3}$ contribution. The scheme  implemented here is entirely different: both 
the propagators are equipped with their cutoffs and the corresponding scales are sent to zero "in parallel". The perturbative procedure can be mimicked by sending the cutoff on fermions ($\Lambda_f$) to zero 
first, which fully dresses the Bose propagator and therefore should recover the standard Landau damping via the last two terms in Eq.~(\ref{eq:pi}). If the limit $\Lambda\to 0$ is considered only subsequently, one anticipates to recover the Hertz-Millis type flow of bosons from the first two terms in Eq.~(\ref{eq:pi}) and also obtain  the non-Fermi liquid from Eq.~(\ref{eq:sigma}). As we discuss in Sec.~III.A.1, this is indeed what happens. The point is that taking the limit $\Lambda_f\to 0$ first leads to the known problem of singular Bose interactions and should therefore be avoided.

\subsection{Regularized fermionic bubbble}

We begin with a discussion of the fermionic contribution to the bosonic self-energy flow as represented by the last two terms of Eq.~(\ref{eq:pi}), the corresponding expression is defined as  
\begin{align}
\mathcal{X}_{0}(Q,\lf):=2g^2\int_K \mathcal{S}_{f, 0}(K)\left(G_{f, 0}(K+Q)+G_{f, 0}(K-Q)\right)\, .
\label{bubble_reg}
\end{align} 
An analogous expression, retaining dressed propagators instead of $G_{f, 0}$ and  $\mathcal{S}_{f, 0}$ will be denoted as $\mathcal{X}(Q,\lf)$. 
A practical calculation requires specifying the form of the fermionic cut-off function, which is taken as \cite{Homenda_2024}: 
\begin{align}
\label{eq:Rf}
R_f(\vec{k})=\begin{cases} 
\big(
    \xi_{k_f+\lf}-\xi_{\vec{k}}
\big)
\theta 
\big[
    \lf-(|\vec{k}|-k_f)
\big]
\; \; {\rm for}\; |\vec{k}|\geq k_f \\ 
\big(
    \xi_{k_f-\lf}-\xi_{\vec{k}}
\big)
\theta 
\big[
    \lf-(k_f-|\vec{k}|)
\big]
\;\; {\rm for}\; |\vec{k}|< k_f\;.
\end{cases}
\end{align} 
Evaluation of the integrals involved in Eq.~(\ref{bubble_reg}) is presented in Appendix B. We also introduce 
\begin{eqnarray}
B\left( Q , \lf \right):=\int_{\l_{UV}}^{\l} d\l' \; (\partial_{\l'} \lf) \mathcal{X}(Q, \lf(\l'))\;,    
\end{eqnarray}
which represents the contribution to the bosonic self-energy from integrating fermions down to the scale $\l$. Note again the presence of two cutoff scales $\Lambda$ and $\Lambda_f$.  The former plays the role of the bosonic cutoff and the flow parameter at the same time. The latter defines the support of the fermionic cutoff function  [see Eq.~(\ref{eq:Rf})]. 
By changing the integration variable we  get rid of explicit dependence on $\l$. At the present truncation level we obtain: 
\begin{eqnarray}
B_0\left( Q , \lf \right):=\int_{\l_{UV,f}}^{\lf} d\l' \; \mathcal{X}_0(Q, \l')\;.    
\end{eqnarray}
The precise form of the relation $\Lambda_f(\Lambda)$ is not specified at this point and will be discussed in the following sections, while an improved truncation involving fermionic self-energy feedback is presented in section IV.
The essential improvement over the Hertz-Millis theory,  proposed in Ref.~\cite{Homenda_2024} amounts to using 
\begin{eqnarray}
\label{Gb}
    G_b(Q, \lf) := \frac{1}{m_b^2 + |\vec{q}|^2 + q_0^2 + B_0(Q, \lf)} 
\end{eqnarray}
for the bosonic propagator instead of the fully-dressed Hertz-Millis form. We refer to the bosonic action involving contributions
from fermions described by $B_0(Q, \lf)$ as the generalized Hertz action \cite{Homenda_2024}. 
% The analytical form of $B_0(Q, \lf)$ is given in the Appendix - see Eq. \dots. 

\begin{figure}
\includegraphics[width = \linewidth]{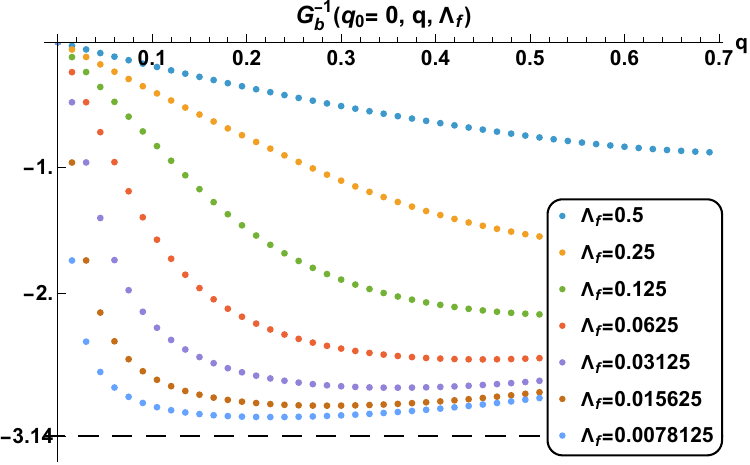}
\caption{The inverse propagator $G_b^{-1}$ given by Eq.~(\ref{Gb}), plotted as a function of $|\vec{q}|$ for $q_0=0$ and a sequence of values of $\Lambda_f$. The minimum defining the ordering wavevector $Q_\Lambda$ approaches zero for $\Lambda_f\to 0$. The  convergence of $G_b^{-1}$ is non-uniform in this limit. The dashed line marks the ultimate bosonic mass shift $\lim_{\Lambda \to 0}\Delta m_b^2 = - \pi \mathcal{A} = -g^2 k_f /(\pi v_f)$. In this calculation we set the initial bosonic mass $m_b^2 = 0$ and $\mathcal{A} = 1$.}  
\label{fig:Gb_sequence}
\end{figure}

For every $\Lambda>0$ the function $G_b^{-1}$ exhibits a minimum at $q_0=0$ and $|\vec{q}|=Q_\Lambda>0$, such that the ordering wavevector remains finite and scales to zero only for $\lf \to 0$. A sequence of plots of $G_b^{-1}$ for different values of $\lf$ is presented in Fig.~\ref{fig:Gb_sequence}.

We also observe that $\mathcal{X}_{0}(Q,\lf)$ is identically zero for $|\vec{q}|=0$ such that no mass flow from fermions arises  if one attempts to artificially constrain the ordering wavevector to zero, which is an unwelcome property encountered and discussed in Ref.~\cite{Fitzpatrick_2013}. This  feature is absent in our present framework, where ${Q}_\Lambda$ is a flowing quantity. %\pawel{Since the flow of bosonic mass is fully controlled by the fermionic regulator and the effective dimension $>d_{upp} = 4$, we are not forced to introduce additional bosonic cut-off function.}   \mateusz{Since the flow of bosonic mass is fully controlled by the fermionic regulator, we are not forced to introduce additional bosonic cut-off function.}

\subsubsection{Recovering the Hertz-Millis theory}

As demonstrated in Ref.~\cite{Homenda_2024}, the Hertz-Millis framework can be recovered from our present theory in a questionable procedure of sending $\Lambda_f$ to zero much faster than $\Lambda$. This can be realized for example by imposing the following relation $\Lambda_f(\Lambda)$:
\begin{equation}
\Lambda_f =(\Lambda-\Lambda_0)\theta (\Lambda-\Lambda_0)   
\label{cutoff_HM}
\end{equation}
with $\Lambda_0>0$, such that $\Lambda_f$ becomes zero for a positive value of $\Lambda$. The standard Landau damping term is then generated from a contribution to the asymptotic form of $B_0(Q, \Lambda_f)$ valid for $|\vec{q}|\gg\Lambda_f$, which has the form: 
\begin{align} 
\label{B_w}
B_> :=  \mathcal{A} \frac{q_0}{v_f |\vec{q}|}\left[\arctan\frac{2v_f|\vec{q}|}{q_0}-\arctan\frac{2v_f\Lambda_f}{q_0}\right]\;,    
\end{align}
where  $\mathcal{A}=g^2 k_f/(\pi^2 v_f)$, for the derivation see the Supplemental material to Ref.~\cite{Homenda_2024} or Appendix C in the present manuscript. By taking $\Lambda_f\to 0$ and considering the static limit $|q_0|\ll v_f|\vec{q}|$, ignoring the generated singular bosonic interactions,  one recovers a generalization \cite{Jakubczyk_2008, Bauer_2011} of the Hertz-Millis framework, in which the RG flow is governed exclusively by bosonic fluctuations. The parameter $\Lambda_0$ may then be interpreted as the upper cutoff of the bosonic effective theory.  

It is clear that the above procedure relying on Eq.~(\ref{cutoff_HM}) and leading to the Landau-damping term is theoretically unsatisfactory. In fact,  the structure of the  coupled Fermi-Bose theory developed below is entirely distinct, in particular it involves a different form of the Bose propagator and contributions to the loop integrals from momenta with $|\vec{q}|<\Lambda_f$, where the Hertz-Millis parametrization is invalid. 

\subsubsection{Flowing ordering wavevector}

We now analyze the structure of the regularized Fermi bubble for $q_0=0$. In the scaling limit, where $max(\Lambda,\Lambda_f, q)\ll k_f$, we obtain 
\begin{eqnarray}
B_0(0,q,\lf)=\mathcal{A} \int_{\Lambda_{f,UV}}^{\Lambda_f} {\rm d} \lambda  \frac{h(\lambda/q)}{q} = 
    \mathcal{A} \int_{\Lambda_{f,UV}/q}^{\Lambda_f/q} {\rm d} t h(t)\,.   
\end{eqnarray} 
Here we denote  $|\vec{q}|=q$ and the function $h(t)$ is given in the Appendix B.2 and plotted in Fig.~\ref{fig:h_function}
%, while $\mathcal{A}= \tfrac{g^2 k_f}{\pi^2 v_f}$ 
. We concentrate on the scaling regime, where $\Lambda_{f,UV}$ is  sent to infinity. 

\begin{figure}
\includegraphics[width = 1.\linewidth]{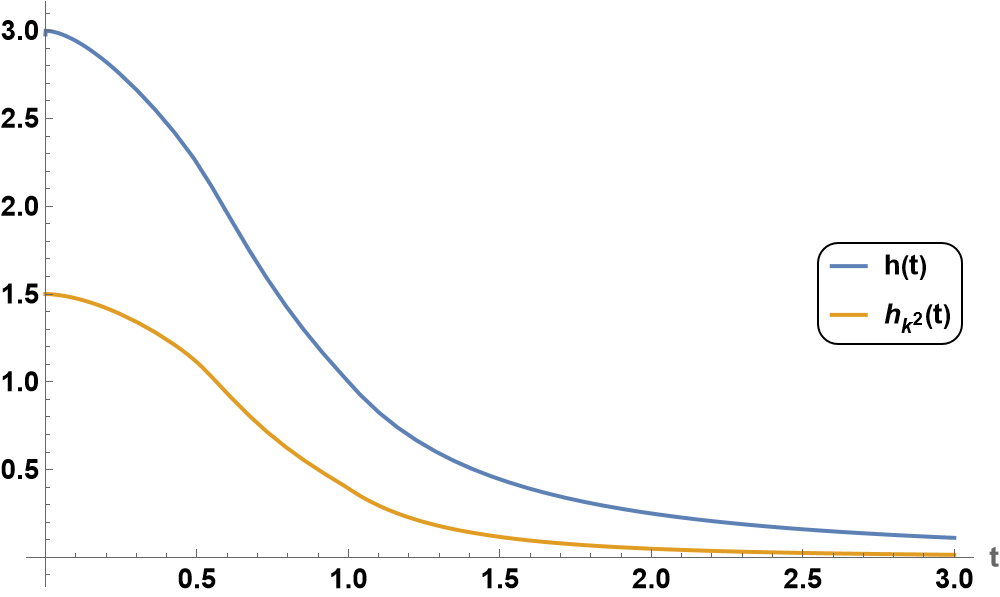}
\caption{Plot of the function $h(t)$ and the function $h_{k^2}(t)$ (for the latter see Sec.~\ref{sec:vf} and Appendix \ref{sec:appendix:scaling} ). } 
\label{fig:h_function}
\end{figure}

\begin{figure}
\includegraphics[width = 1.\linewidth]{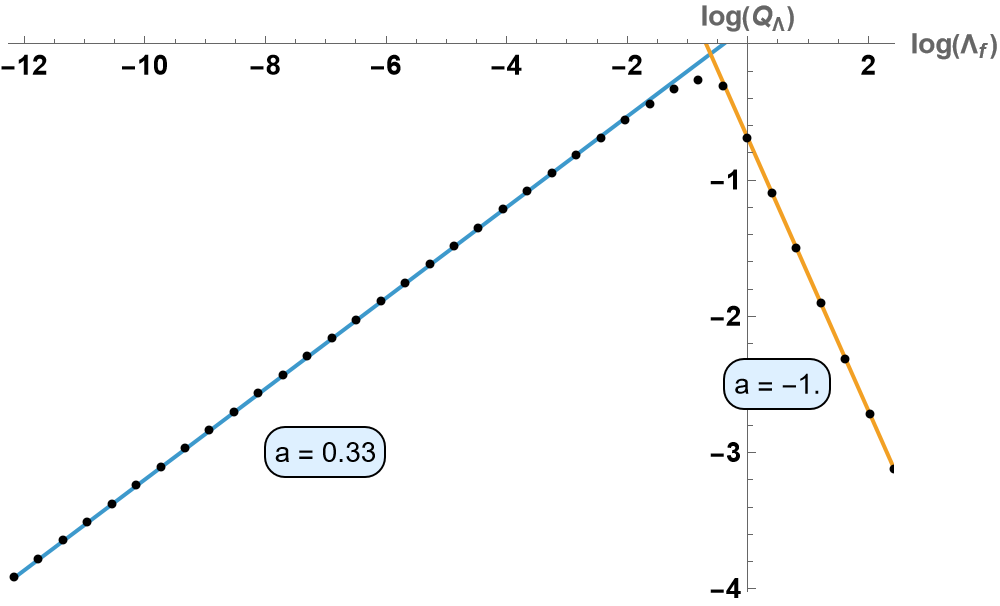}
\caption{Comparison of $Q_\Lambda$ obtained numerically from Eq.~(\ref{Q_eq}) (black points) and the asymptotic expressions of Eq.~(\ref{Q_Lambda_asympt}) (solid lines). The numbers indicate slope fits to the numerical data. We set $\mathcal{A} = 1$. } 
\label{fig:Q_Lambda}
\end{figure}

We identify the scale-dependent ordering wavevector $Q_\Lambda$ by searching for minima of $B_0(0,q,\lf)+q^2$ with respect to $q$, which leads to 
\begin{eqnarray} 
\label{Q_eq}
    2 Q_\Lambda - \mathcal{A} \; \frac{\Lambda_f}{Q_\Lambda^2}h(\frac{\Lambda_f}{Q_\Lambda})
    = 0\;.   
\end{eqnarray} 
Eq.~(\ref{Q_eq}) can be analytically solved in the asymptotic regimes of $\Lambda_f\gg Q_\Lambda$ and $\Lambda_f\ll Q_\Lambda$, which yields 
\begin{eqnarray} 
\label{Q_Lambda_asympt}
Q_\Lambda\approx \frac{\mathcal{A}}{2\Lambda_f}\;,\;\;\;{\rm and}\;\;\; Q_\Lambda\approx \left(\frac{3\mathcal{A}}{2}\Lambda_f\right)^{1/3}    
\end{eqnarray}
respectively. We are however presently interested only in the limit where both $\Lambda_f$ and $Q_\Lambda$ vanish, which corresponds to the latter of the two above solutions. Comparison of the obtained asymptotic expressions with direct numerical evaluation of $Q_\Lambda$ is given in Fig.~\ref{fig:Q_Lambda}.  

We point out that resolving the full momentum profile of the function $B_0(0,q,\lf)$ is necessary for obtaining the correct scaling of $Q_\Lambda$. Neglecting analytical terms in $\mathcal{X}_0$ as was proposed in Ref.~\cite{Homenda_2024} leads to $Q_\Lambda\sim \Lambda_f$ instead of $Q_\Lambda\sim \Lambda_f^{1/3}$. The  obtained result for $Q_\Lambda$ implies that $\epsilon_\Lambda:=\Lambda_f/Q_\Lambda$ may be considered as a small parameter. This offers an avenue for simplifications relying on expansion of $B_0$ in $\epsilon_\Lambda$, as an alternative to the analysis of Ref.~\cite{Homenda_2024} which used the full form  $B_0(Q,\lf)$.

We proceed by evaluating the contribution to the flow of the boson  mass from fermions. 
%Taking $\Lambda_{f, UV}\to \infty$ 
We obtain: 
\begin{eqnarray} 
\label{rob}
B_0(0,Q_\Lambda,\lf)\approx \Delta m_{b,0}^2+2Q_\Lambda^2   \;, 
\end{eqnarray}
valid for $Q_\Lambda\gg \Lambda_f$, see Appendix C. The constant mass shift is given as $\Delta m_{b,0}^2= - \pi \mathcal{A}$. The present result  suggests that at RG fixed point  $\Lambda_f\sim\Lambda^3$, which is required to obtain the same scaling of the two terms in Eq.~(\ref{rob}) and the conventional scaling $m_b^2\sim\Lambda^2$ of the boson mass. We elaborate further on this point below.

\subsubsection{Frequency dependence and Landau damping} 

We are now ready to explore the frequency dependence of the flowing fermion bubble diagram. One of our aims is to extract the exponent $z_b$ determined by the characteristic frequency $q_0\sim \Lambda^{z_b}$. We evaluate $B_0$ at the flowing ordering wavevector $Q_\Lambda$. The corresponding integrand can, in analogy to the static case $q_0=0$, be expressed as
\begin{eqnarray} \label{eq:NSC:scalingCHI}
    \mathcal{X}_{0}(q_0,q, \lf) =  
    \frac{ \mathcal{A}}{q} \;
    h\bigg( \frac{\Lambda_f}{q}, \frac{1}{v_f} \frac{q_0}{q}\bigg) 
\end{eqnarray}
--- see Appendix B. We introduce 
\begin{eqnarray}
    w=\frac{q_0}{v_f Q_\Lambda}
\end{eqnarray}
and obtain: 
\begin{eqnarray}
\label{B000}
    &B_0(q_0, Q_\Lambda, \lf) = 
    & \mathcal{A} \left[ 
    \int_{\infty}^{\frac{1}{2}}{\rm d} t \,  
    h \big(t, w \big)
    + \int_{\frac{1}{2}}^{\frac{\Lambda_f}{Q_\Lambda}}{\rm d} t \, 
    h \big(t, w  \big)
    \right] \, .
\end{eqnarray}
The $h(t)$ function used previously corresponds to the static limit $h(t, w=0)$. The first term is expandable in $w$ and yields, for $w\ll 1$, the leading contribution of order $w^2$. Terms of a different type arise from the vicinity of the upper integration limit in the second term, which is of order $\epsilon_\Lambda$ (see  Appendix C for details). Note that, by virtue of the result of the previous section, we have $\epsilon_\Lambda=\Lambda_f/Q_\Lambda\ll  1$.  We obtain:  
\begin{eqnarray} \label{eq:NSC:Bdynamic}
    B_0(q_0, Q_\Lambda, \lf)
    = - \mathcal{A} \Bigg[
    \frac{3}{2} - \frac{3 \Lambda_f}{Q_\Lambda} + 
    \frac{1}{2}\frac{q_0^2}{q_0^2 + v_f^2 Q_\Lambda^2} - \notag \\
        \frac{\Lambda_f}{Q_\Lambda}\frac{q_0^2}{q_0^2 + 4 v_f^2 \Lambda_f^2} +
        \frac{2 q_0}{v_f Q_\Lambda}\bigg( \arctan\big(\frac{2 v_f \Lambda_f}{q_0}\big) \notag \\ 
    - \arctan\big(\frac{ v_f Q_\Lambda}{q_0}\big) \bigg) 
    \Bigg] 
    +  \mathcal{O}\left( w^2\right) \; . 
\end{eqnarray}

\begin{figure}
    \centering
    \includegraphics[width=1.\linewidth]{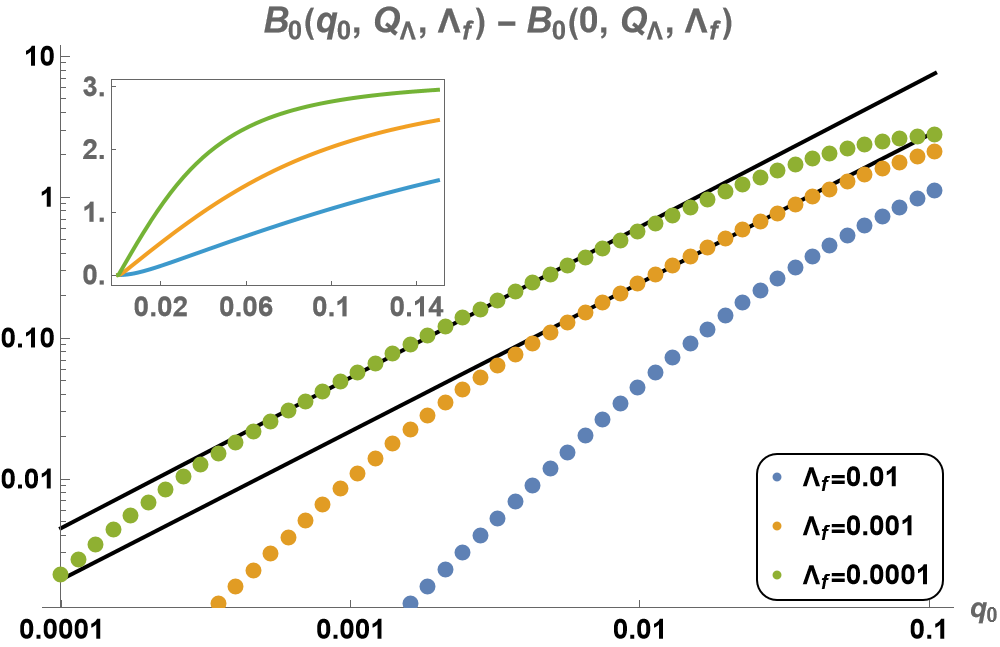}
    \caption{Dynamical part of the function $B_0(q_0, q, \Lambda_f)$. Linear scaling is clearly visible in the  frequency range between $2v_f \Lambda_f$ and $v_f Q_\Lambda$ (see the main text). Quadratic dependence on frequency sets in for $q_0<2v_f \Lambda_f$. The inset shows the same data in linear scale. }
    \label{fig:dynamicalB}
\end{figure}

The range of variation of the above expression, treated as a function of $q_0$ spans the interval $q_0\in (2v_f\Lambda_f,v_f Q_\Lambda)$, which contains frequencies for which  $ w\sim Q_\Lambda^2$. In this regime we expand the first $\arctan(x)$ function around $x=0$ and the second one around infinity. This leads to a contribution of order $|w|$ and yields a simple parametrization of the flowing bubble and the bosonic propagator: 
\begin{eqnarray}
    &&B_0( q_0, Q_\Lambda, \lf) \approx 
    - \frac{3 \mathcal{A}}{2} + 2 Q_\Lambda^2 + \mathcal{A}\, \pi \frac{|q_0|}{v_f Q_\Lambda}\;, \\
    &&G_{b} \approx \frac{1}{\Delta m_b^2 + 3 Q_\Lambda^2 + (|\vec{q}| - Q_\Lambda)^2 + \mathcal{A}\pi \frac{|q_0|}{v_f Q_\Lambda}} \; . 
    \label{G0p}
\end{eqnarray} 
For an illustration see Fig.~\ref{fig:dynamicalB}.
We note affinity of the above expression to the Hertz-Millis form pertinent to the {\it antiferromagnetic}-type transition where the ordering wavevector $\vec{Q}\neq \vec{0}$ and the (one-loop) dynamical exponent of the Bose field $z_b=2$. However, taking note of the analysis of the previous section and requiring that the distinct terms in $G_b^{-1}$ all scale as $\Lambda^2$, we obtain 
\begin{eqnarray}
    |q_0| \sim Q_\Lambda^3 \sim \Lambda_f  \;,
\end{eqnarray}
which leads to $z_b=3$, in agreement with the Hertz-Millis theory. Note that, in spite of the very  different forms of the  Bose  propagators, the same value of $z_b$ is obtained in our approach if fermions are scaled out much faster than bosons  - see Sec.~III.A.1, resulting in the standard Hertz-Millis propagator instead of the form of Eq.~(\ref{G0p}). 

The key point of the above analysis is the identification of the relation $Q_\Lambda\sim\Lambda\sim\Lambda_f^{1/3}$, establishing the relative rate at which the distinct degrees of freedom need to be scaled. The earlier  reasoning of Ref.~\cite{Homenda_2024} 
disregarded analytical contributions to the regularized static part of the bubble, which led to the flawed result $\Lambda_f\sim\Lambda\sim Q_\Lambda$ and $z_b=2$. 
The other point to note is that the expanded form of Eq.~(\ref{G0p}) is valid only for small $w$, but the complete formula for $B_0(Q, \lf)$, derived and implemented in Ref.~\cite{Homenda_2024} necessarily spanned the entire range of momenta and frequencies, although it neglected analitycal contributions in $w$.
% applies to the entire range of momenta and frequencies (see Sec.~III.B for further discussion).   

We emphasize once again that the above calculation entirely evades the step of integrating the fermionic degrees of freedom before the bosons and constitutes a robust   starting point for further developments accounting, in particular, for self-energy effects --- see Sec.~IV.

\subsection{Fermion self-energy} 
\begin{figure*}
{\includegraphics[width=0.48\linewidth]{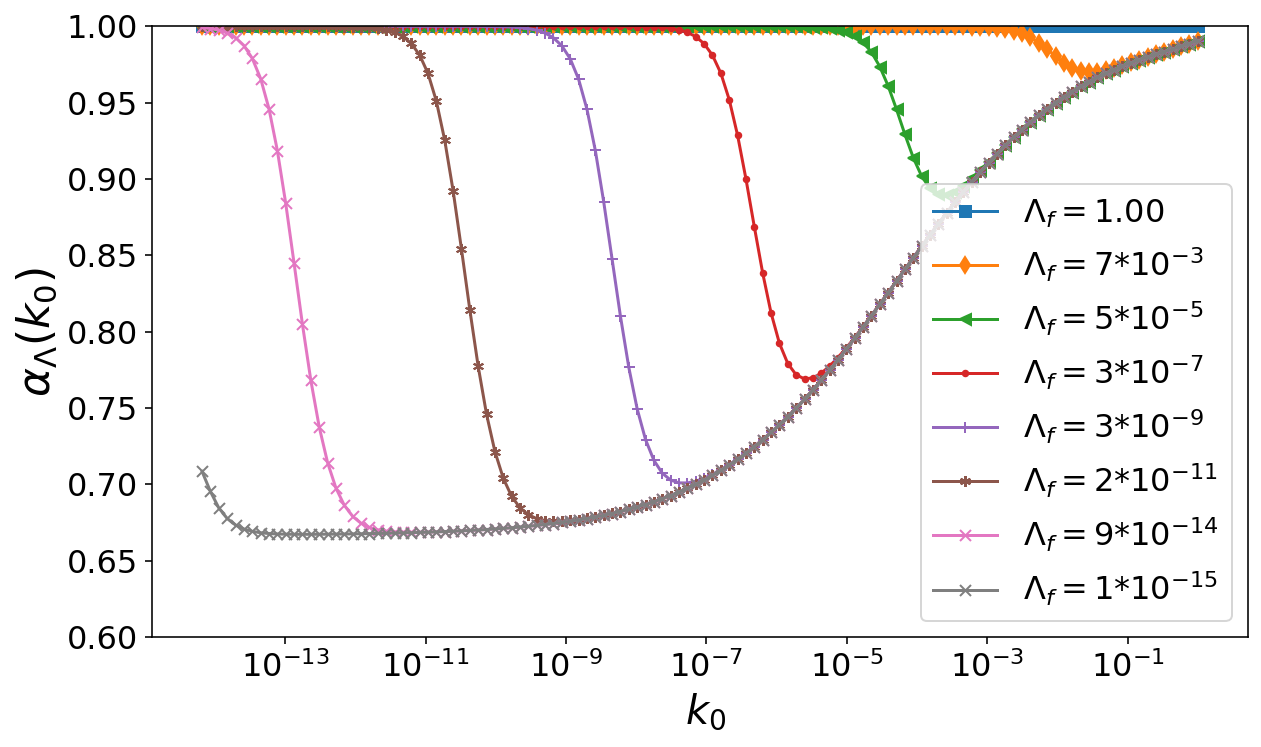}} 
{\includegraphics[width=0.48\linewidth]{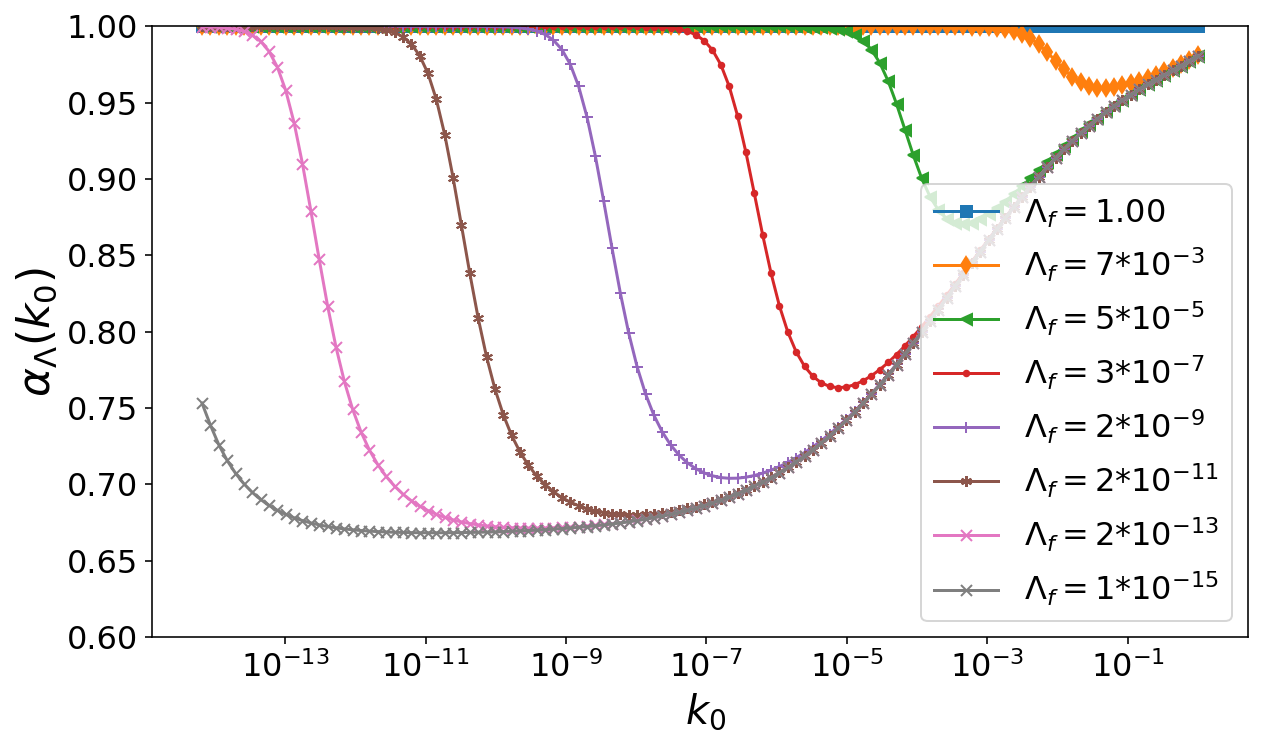}}
    \caption{Evolution of $\alpha_\Lambda (k_0)$ upon reducing the cutoff scale $\Lambda_f$. In the calculation presented in the left panel we disregarded the term involving $\partial_\Lambda R_b$. In both cases we reach the anticipated non-Fermi liquid scaling with $\alpha \approx 2/3$. For
low frequencies and finite $\Lambda$ we observe the crossover back to the Fermi liquid behavior. The frequency $\tilde{k}_0(\lf)$, identified with this crossover, is plotted in Fig.~\ref{fig:freq:cros}. The initial boson mass was tuned so that the system reaches quantum critical scaling.}
    \label{fig:freq:alpha} 
\end{figure*}

\begin{figure}
    \centering
\includegraphics[width=\linewidth]{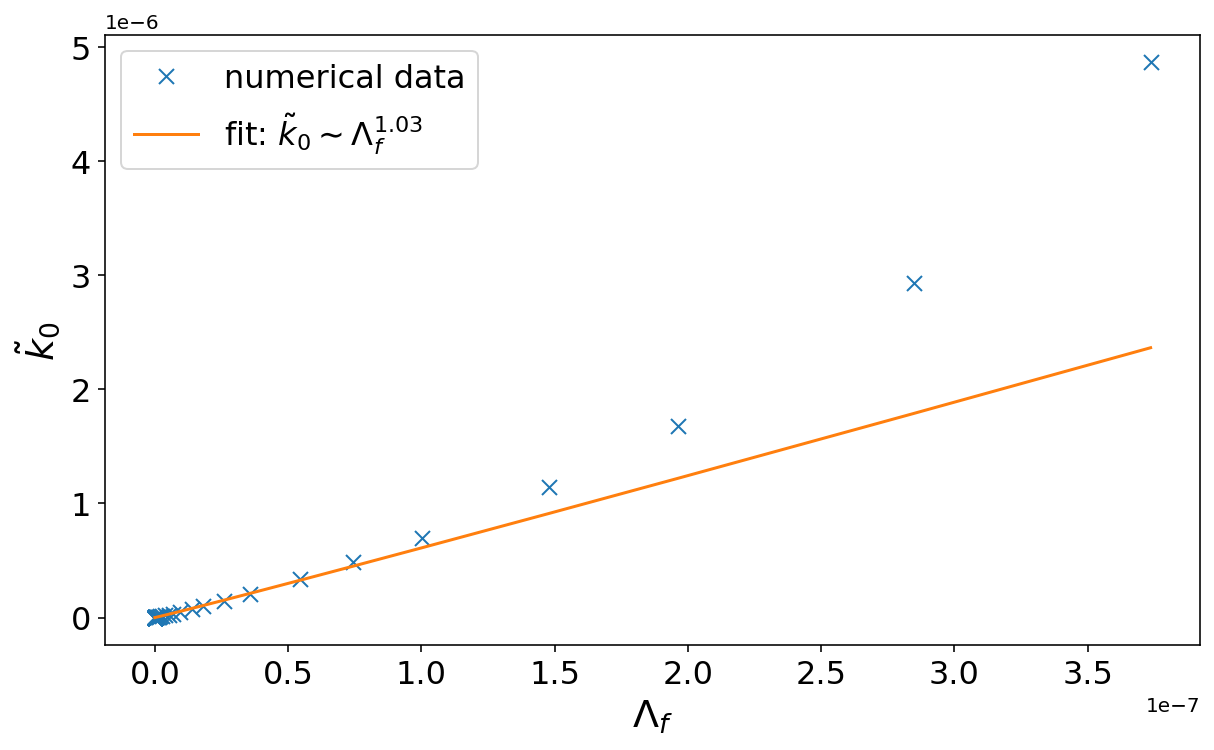}
    \caption{The crossover frequency estimated via  $\alpha_{\l}(k_0) = \alpha^* + 0.1 \approx 0.77$ ($\alpha^*$ is a fixed point value). There is no visible difference between the two variants of the calculation.  
    We deduce that $\tilde{k}_0 \sim \Lambda_f \sim \l^3$.}
    \label{fig:freq:cros}
\end{figure}

\begin{figure}
    \centering
    \includegraphics[width=\linewidth]{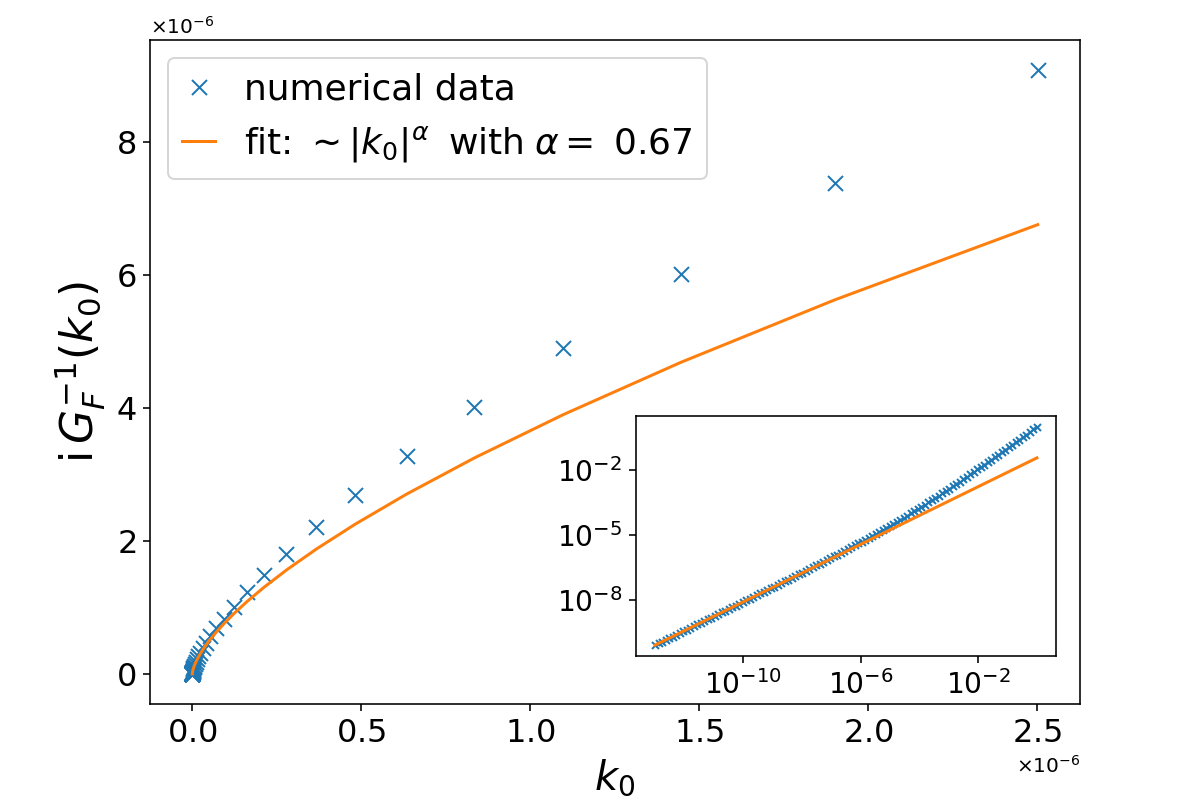}
    \caption{Frequency dependence of $G_f^{-1}$ from integrating the flow down to $\Lambda_f \approx 0$. At low frequencies we recover the RPA-type scaling $\sim k_0^{2/3}$. The inset shows the same data in logarithmic scale. There is no visible difference between the two variants of the calculation.}
    \label{fig:freq:Ginv}
\end{figure}

We now explain how the one-loop result for the Fermi self-energy and the associated non-Fermi liquid state is recovered in our approach relying on the generalized Hertz action of the previous sections. We define: 
\begin{align}
    \Sigma(k_0) \coloneqq \Sigma(k_0, \vec{k} = k_f \hat{e}_x) \; 
\end{align}
and concentrate for now on the imaginary part: 
\begin{align}
    \tilde{\Sigma}(k_0) \coloneqq k_0 - {\rm Im} \left[ \Sigma(k_0) \right] \; .
\end{align}
The non-Fermi liquid propagator is generated  from the  Fermi-Bose loop [Eq.~(\ref{eq:sigma})], which reduces to the   form: 
\begin{align} \label{eq:freq:sigmaflow}
    \partial_\Lambda \tilde{\Sigma}(k_0) = 2 g^2 \int_Q 
    \frac{ (\partial_\Lambda \Lambda_f) \, \partial_{\Lambda_f} R_f(k_f \hat{e}_x + \vec{q})}{m_b^2 + |\vec{q}|^2 + B_0(q_0, q, \lf) + R_b(|\vec{q}|)} \nonumber \\
    \frac{(q_0 - k_0) \, f(k_f \hat{e}_x + \vec{q})}{\left[ (q_0 - k_0)^2 + \big(f(k_f \hat{e}_x + \vec{q}) \big)^2 \right]^2} \; \nonumber \\
    + g^2 \int_Q 
    \frac{\partial_\Lambda R_b(| \vec{q}|)}{\big(m_b^2 + \Lambda_b^2 + B_0(q_0, q, \lf) \big)^2} \nonumber \\
    \frac{q_0 - k_0 }{ (q_0 - k_0)^2 + \big(f(k_f \hat{e}_x + \vec{q}) \big)^2 } \; ,
\end{align} 
where we introduced $f(\vec{k})=\xi_{\vec{k}}+R_f(\vec{k})$.
The term involving $\partial_\lf R_f$ is crucial, dropping the contribution with $\partial_\l R_b$ has no important impact on the results of this section.  
The bosonic propagator introduces a characteristic scale $\tilde{q}_0 \sim \Lambda^{z_b}$, effectively acting as a cut-off for frequencies above $\tilde{q}_0$. 
In the analysis to follow, we pursue two strategies. In the first calculation, we directly numerically integrate Eq.~(\ref{eq:freq:sigmaflow}) using bosonic propagator given in Eq.~(\ref{G0p}), which yields the entire profile of $\tilde{\Sigma}(k_0)$ including all the relevant regimes both as function of frequency $k_0$ and cutoff scale $\lf$ (one can use $\l$ equivalently). We define the flowing exponent:
\begin{align}
    \alpha_\Lambda (k_0) \coloneqq \pdv{\ln (\tilde{\Sigma}(k_0))}{\ln (k_0)} \; ,
\end{align} 
which encodes the scale dependence of the function. The corresponding results demonstrating the flow of $\alpha_\l(k_0)$  are given in Fig.~\ref{fig:freq:alpha}. 
Due to the presence of the cutoff at finite $\l$, the bosonic field acquires non-zero mass $m_{b}^2 \sim \l^2$, which suppresses the non-Fermi liquid scaling and restores the Fermi liquid in deep infrared as long as $\l>0$. Directly at the QCP, the frequency range of this scaling will shrink and ultimately vanish as $\l$ is reduces towards zero. On the other hand, if the system becomes detuned from criticality the bosonic mass would freeze at a certain scale $\tilde{\l}$ and stay finite till the end of the flow. We clearly see that for  $k_0>\tilde{k}_0$ we recover the anticipated scaling of the self-energy governed by the exponent $\alpha\approx 2/3$, while for $k_0<\tilde{k}_0$ the self-energy crosses back to the Fermi liquid behavior ($\alpha =1$). In the UV regime we recover linear dependence of $\Sigma$ on frequencies. The scale dependence of the  crossover frequency and the frequency profile of $G_f^{-1}$ at the QCP are presented in Figs.~\ref{fig:freq:cros} and ~\ref{fig:freq:Ginv}. The crossover frequency coincides with the frequency scale imposed by a bosonic propagator $\tilde{k}_0 \sim \lf \sim \l^3 \sim \tilde{q}_0$.


In the other, simplified variant of the calculation, we take advantage of the fact that for $k_0 \ll \tilde{q}_0$, we can expand the RHS of Eq.~(\ref{eq:freq:sigmaflow}) in powers of $k_0$. As a result we find that $\partial_\Lambda \tilde{\Sigma}(k_0) \sim k_0$. This allows us to adopt a commonly applied parametrization:
\begin{align}
    \tilde{\Sigma}(k_0) \approx Z_f k_0 \quad {\rm for} \quad k_0 < \Lambda^{z_b} \; ,
\end{align}
where $Z_f$ depends on the scale $\lf$.
%For suficiently large fermionic frequencies $k_0 \approx \tilde{q}_0$
In this regime, the frequency dependence of the Fermi self-energy can be entirely determined from its low-energy expansion:
\begin{align}
\label{eq:freq:sigmaflowSimpl}
    \partial_\Lambda \tilde{\Sigma}(k_0) = 2 g^2 \int_Q 
    \frac{ (\partial_\Lambda \Lambda_f) \, \partial_{\Lambda_f} R_f(k_f \hat{e}_x + \vec{q})}{m_b^2 + |\vec{q}|^2 + B(q_0, q, \lf)} \nonumber \\ 
    \frac{ Z_f \;(q_0 - k_0) \, f(k_f \hat{e}_x + \vec{q})}{\left[  Z_f^2 \; (q_0 - k_0)^2 + \big(f(k_f \hat{e}_x + \vec{q}) \big)^2 \right]^2} \nonumber \\
    + g^2 \int_Q 
    \frac{\partial_\Lambda R_b(| \vec{q}|)}{\big(m_b^2 + \Lambda_b^2 + B_0(q_0, q, \lf) \big)^2} \nonumber \\
    \frac{Z_f(q_0 - k_0) }{ Z_f^2 (q_0 - k_0)^2 + \big(f(k_f \hat{e}_x + \vec{q}) \big)^2 } \; .
\end{align} 
Note that in this approach we may easily include the self energy in the loop, which amounts to multiplying the frequencies by $Z_f$ (as done in the above equation). 
The flow of $Z_f$ is extracted from Eq.~(\ref{eq:freq:sigmaflowSimpl}) by taking the derivative with respect to frequency at $k_0 = 0$, which leads to:
\begin{align} \label{eq:freq:flowZf}
    \partial_\Lambda & Z_f = - 2 g^2 Z_f
    \int_Q 
    \frac{ (\partial_\Lambda \Lambda_f) \, \partial_{\Lambda_f} R_f(k_f \hat{e}_x + \vec{q})}{m_b^2 + |\vec{q}|^2 + B(q_0, q, \lf)} \nonumber \\ \nonumber
     &\frac{ f(k_f \hat{e}_x + \vec{q})}{\left[ Z_f^2 q_0^2 + \big(f(k_f \hat{e}_x + \vec{q}) \big)^2 \right]^2}
    \left(
    1 - 
    \frac{4 Z_f^2 q_0^2}{ Z_f^2 q_0^2 + \big(f(k_f \hat{e}_x + \vec{q}) \big)^2 }
    \right) + \\ \nonumber
    & \qquad - g^2 Z_f
    \int_Q 
    \frac{\partial_\Lambda R_b(| \vec{q}|)}{\big(m_b^2 + \Lambda_b^2 + B_0(q_0, q, \lf) \big)^2} \\ 
     &\frac{1}{ Z_f^2 q_0^2 + \big(f(k_f \hat{e}_x + \vec{q}) \big)^2 }
    \left(
    1 - 
    \frac{2 Z_f^2 q_0^2}{ Z_f^2 q_0^2 + \big(f(k_f \hat{e}_x + \vec{q}) \big)^2 }
    \right) \; .     
\end{align}
 
We then examine the power counting dictating the scaling of $Z_f$. On the RHS of Eq.~(\ref{eq:freq:flowZf}) the bosonic propagator scales as $G_b \sim Q_\Lambda^{-2} \sim \Lambda_f^{-2/3}$, and the fermionic part as $\sim v_f^{-4} \Lambda_f^{-4} $. The frequency integral yields an additional factor $v_f \Lambda_f/Z_f$ , while the momentum integrals contributes $Q_\Lambda \Lambda_f \sim \Lambda_f^{4/3}$.  In total we obtain: 
\begin{align} \label{eq:freq:ZfNSC}
    \partial_{\Lambda_f} Z_f &\sim - 
    \frac{g^2 Z_f v_f^2 \Lambda_f}{2 \pi^3}
    \underbrace{\frac{ v_f \Lambda_f}{Z_f}}_{{\rm d}q_0}
    \underbrace{\frac{1}{v_f^4 \Lambda_f^4}}_{fermionic} \;
    \underbrace{\frac{1}{\Lambda_f^{2/3}}}_{bosonic} 
    \underbrace{\Lambda_f^{4/3}}_{{\rm d}q^2} \notag \\
    \;  &= -  \frac{g^2 }{v_f \Lambda_f^{4/3} } 
    \quad
    \xRightarrow{\int d \Lambda_f} 
    \quad
    \eta_f = \frac{1}{3}\;,
\end{align} 
where the exponent $\eta_f$ is defined by 
\begin{equation}
    \eta_f:=-\frac{1}{Z_f}\frac{d Z_f}{d\ln\Lambda_f}\;. 
\end{equation}

 Interestingly $Z_f$  cancels out in  Eq.~(\ref{eq:freq:ZfNSC}), which suggests that the result $\eta_f = 1/3$ may  remain unchanged no matter if we retain or neglect the fermionic self-energy on the RHS of the loop integral. This is reminiscent of the behavior obtained at RPA \cite{Dellanna_2006}, where results from the self-consistent and non self-consistent variants of the analysis yield the same behavior of the self-energy. 
 
The relation between $\alpha$ and $\eta_f$ can be derived from  continuity of $\tilde{\Sigma}(k_0)$ at $k_0 =  \tilde{k}_0$. Within the present parametrization, for small enough frequencies the fermionic propagator can be written as:
\begin{eqnarray}
    \tilde{\Sigma}(k_0 < \tilde{k}_0) \sim \lf^{-\eta_f} k_0 \xrightarrow{k_0 \to \tilde{k}_0^-} \lf^{- \eta_f} \lf \; ,
\end{eqnarray}
while for frequencies from the IR non-Fermi liquid region we get :
\begin{eqnarray}
    \tilde{\Sigma}(\tilde{k}_0< k_0 < k_0^{UV}) \sim k_0^{\alpha} \xrightarrow{k_0 \to \tilde{k}_0^+}  (\lf)^{\alpha} \; .
\end{eqnarray}
Comparing the two limits%from above and below $\tilde{k}_0$
, we obtain the relation:
\begin{eqnarray} \label{eq:alpha_eta}
    \alpha &=& 1 - \eta_f \; ,
\end{eqnarray}
thus leading to the RPA value $\alpha=2/3$ at the present approximation level.

The present section demonstrated how the  framework of the Wetterich equation  recovers the previously established one-loop results for bosonic and fermionic properties at the $\vec{Q}=\vec{0}$ quantum critical point of itinerant Fermi systems.  We contrasted the flawed  procedure (similar in spirit to the Hertz-Millis theory) of scaling $\Lambda_f$ to zero before the bosonic cutoff $\Lambda$ (Sec.~III.A.1) with the more satisfactory one, where fermions and bosons are  integrated out in parallel leading, for the bosonic sector, to the generalized Hertz action. The obtained fixed point governing the critical singularities is of Gaussian nature and leads, after correct account of the scaling of the flowing  ordering wavevector $\vec{Q}_\Lambda$, to the value $z_b=3$ for the dynamical exponent of the Bose field. It also  recovers the 
RPA result for the Fermi self-energy exponent $\alpha=2/3$. 

We emphasize that in spite of the fact that, at this approximation level, the two paths lead to the same values of the scaling exponents, the latter one evades the problem of generating singular effective Bose interactions and the singular Landau-damping term. In the sections to follow, we address the question of stability of the fixed point obtained above, upon taking account of the scaling of the Yukawa coupling $g$ and inclusion of self-energy into the Fermi loop governing the flow of the Bose propagator. 

Before proceeding we briefly comment on the numerical procedures implemented in this  section. We evaluate $\tilde{\Sigma}(k_0)$ on a logarithmic frequency grid with 120 points. The resulting set of flow equations is integrated using an adaptive version of the Dormand-Price method. The internal momentum and frequency integrals are calculated numerically at each time step for each of the external frequencies $k_0$ using the Gauss-Legendre quadrature. If not specified otherwise, the flows exhibited in the paper correspond to the following numerical values of the parameters: ${k_f = 10,~ v_f = 1,
~ g = 2,~ \l_{UV} = 1, \l_{f, 0} \approx 10^{-15}}$. Further analysis of the momentum integrals in Eq.~(\ref{eq:freq:sigmaflow}) is given in Appendix D. Technical details are further discussed in Appendix E.

\section{Self-consistent treatment}
The following discussion concerns stability of the non-Fermi liquid RG fixed point of Sec.~III upon elevating the truncation level. It is performed in two steps. We first include (Sec.~IV.A) scaling of the Yukawa coupling $g$ and the imaginary part of the self energy $\tilde{\Sigma}(k_0)$ in the bubble diagram governing the flow of the boson propagator. We obtain a  result indicating instability of the RPA-type non-Fermi liquid and no possibility of a sensible alternative RG fixed point. 
In the second step (Sec.~IV.B) we account also for the flow of the real part of the self energy, which is parametrized by the flowing Fermi velocity $v_f$. Our results indicate that the non-Fermi liquid RG fixed point is not restored at this truncation level.

One possible simple  explanation of these findings would invoke a generic instability of the $\vec{Q}=\vec{0}$ QCP, which was demonstrated in previous literature for a variety of systems (see e.g. \cite{Chubukov_2004, Belitz_2005,  Kirkpatrick_2026}). In this interpretation, the one-loop fixed point discussed in Sec.~III would appear as a mere approximation artifact. The other possibility, which we discuss in Sec.~IV.C is that the correct account of the presently discussed non-Fermi liquid state requires a significantly higher level of truncation, in particular inclusion of terms absent in the bare effective action of Eq.~(\ref{bare_action}), but generated by the RG flow. We speculate on how inclusion of the vertex of type $\Gamma^{(4)}_{\phi\phi\psi\bar{\psi}}$ might potentially lead to reappearance of a non-Fermi liquid   RG fixed point with $\vec{Q}=\vec{0}$.

\subsection{Yukawa coupling and Fermi self-energy scaling}
The logic of our reasoning is the following: we will assume conventional fixed point scaling of both the fermion self-energy and the Yukawa coupling and subsequently show that this leads to inconsistencies in the flow equations for the self-energies irrespective of the values of the scaling exponents. Our argument does not require analysis of flow equation for $g$.

\begin{figure}[h!]
    \centering  \includegraphics[width=\linewidth]{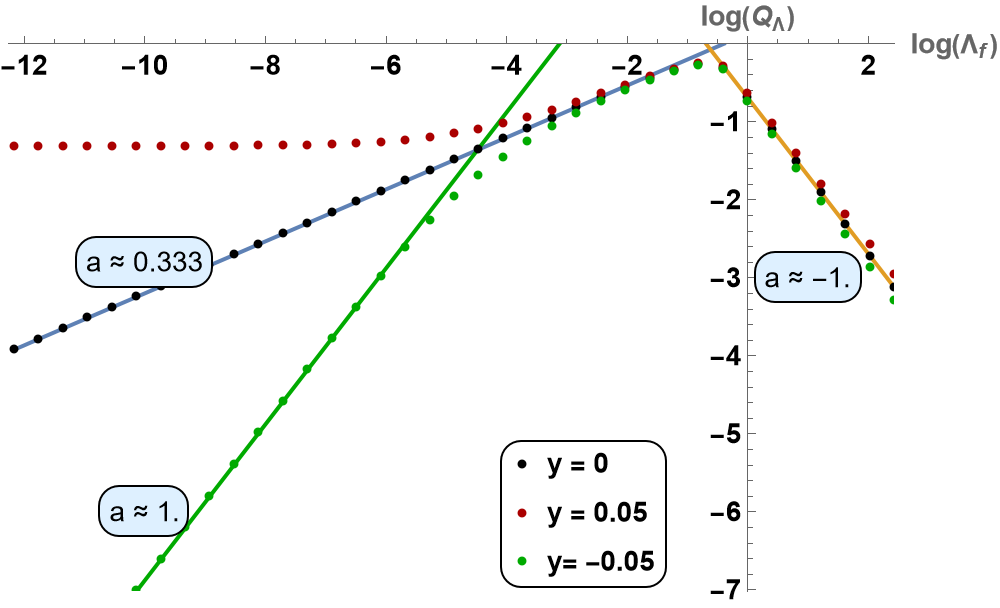} 
\caption{Numerical solution of Eq.(\ref{eq:semi:Qequation}). The numbers indicate slope fits to the numerical
data. In this calculation we set $\mathcal{A}^* = 1$. } \label{fig:semi:Qlnumerical}
\end{figure}

We concentrate on the low-energy regime and reexamine the flow of the self-energies assuming the following fixed-point scaling of $Z_{f, \Lambda}$ and $g_{\Lambda}$:
\vspace{-0.5ex}
\begin{align} \label{parametrization} 
    Z_{f} = Z_f^* \; \Lambda_f^{-\eta_f}\;,\;\;\;\;\; 
    g_{\Lambda} = g^* \; \Lambda_f^{-\eta_g}\;. 
\end{align}
 The regularized fermionic bubble (denoted at this approximation level as $\mathcal{X}_{s}$)  takes the following form:  
\begin{align}
    \mathcal{X}_{s}(Q, \lf) = - 
    \frac{2 g_\Lambda^2}{(2 \pi)^3 Z_{f}} 
    \int_{-\infty}^{\infty} {\rm d} k_0 
    \int {\rm d}^2 k 
    \frac{ \partial_{\lf} R_f(\vec{k}) }{ \big(-i k_0+ f(\vec{k}) \big)^2} \nonumber  \\
    \Bigg[
        \frac{1}{-i (k_0 + Z_f q_0) + f(\vec{k} + \vec{q}) }
         +  
        \frac{1}{-i (k_0 - Z_f q_0) + f(\vec{k} - \vec{q}) }
    \Bigg] \; .
\end{align}
In analogy with the scaling form of Eq.~(\ref{eq:NSC:scalingCHI}), we can rewrite this expression as
\begin{eqnarray} \label{eq:semi:scalingCHI}
    \mathcal{X}_{s}(Q, \lf) = 
    \frac{\mathcal{A}^* \Lambda_f^{y}}{q} \;
    h\bigg( \frac{\Lambda_f}{q}, \frac{Z_{f}^*}{v_f} \frac{q_0}{q^{1 + \eta_f}}\bigg) \; ,
\end{eqnarray}
 where we defined the exponent:
\begin{eqnarray}
    y = - 2 \eta_g + \eta_f \; .
\end{eqnarray}
The  function $h(t, w)$ has the same form as in the approximation discussed in Sec.~III --- see Appendix B. Regarding the static part of the integrated bubble, the modification as compared to the previous case occurs via the $y$ exponent. We obtain:
\begin{eqnarray}
    B_s(0, q,  \lf) = 
\mathcal{A}^*\int_{\Lambda_{f,UV}}^{\Lambda_f} {\rm d} \lambda \;
    \lambda^y \frac{h(\lambda/q)}{q} \;,
\end{eqnarray} 
where $\mathcal{A}^*= \frac{(g^*)^2 k_f}{\pi^2 v_f Z_f^*}$.
The quantity $\mathcal{X}_{s}$ shares all the features of $\mathcal{X}_{0}$ analyzed in the previous section, however the properties of $B_{s}$  now crucially depend on $y$.
Let us examine the equation for the minimum of the inverse bosonic propagator $G_{b, s}^{-1}$:
\begin{eqnarray} \label{eq:semi:Qequation}
    \frac{d}{dq}&&\left( q^2 + B_s(0, q, \lf) \right)_{q = Q_\Lambda}= \notag \\
    && 2 Q_\Lambda + \frac{y}{Q_{\Lambda}} B_{s}(0, Q_{\Lambda}, \lf) - \mathcal{A}^* \; \frac{\Lambda_f^{1+y}}{Q_\Lambda^2} h\Big(\frac{\Lambda_f}{Q_\Lambda}\Big)
    = 0 \; . \quad
\end{eqnarray}
In Fig.~\ref{fig:semi:Qlnumerical}, we present numerical solutions obtained for three exemplary values of $y$: $y \in \{ -0.05, \, 0, \, 0.05 \}$. For sufficiently large $\Lambda_f$, we observe only minor deviations from the solution discussed in Sec.~III. However, for positive $y$,  the flowing order parameter $Q_\Lambda$ saturates for $\Lambda_f$ small. 
This behavior is generic for $y>0$ and indicates instability towards a phase with broken translational symmetry, where $\lim_{\Lambda_f \to 0}Q_{\Lambda} \neq 0$. In contrast, for $y<0$, the scaling of the momentum wavevector deviates from the previously identified asymptotic solution and we obtain $Q_\Lambda\sim \Lambda_f$. This behavior  leads to an inconsistency in the mass flow, which is described below. 
In consequence, $y = 0$ appears at first sight as a necessary condition for occurrence of a second-order phase transition  with $\vec{Q}=\vec{0}$ at this level of truncation. This however yields in turn an unphysical behavior of the self-energy flow  as we discuss in what follows.

To explain the problem, we first analyze the limit:
\begin{eqnarray}
    \lim_{\Lambda \to 0} B_{s}(0,|\vec{q}| = \lf,  \lf) \sim \Delta m_b^2 \; .
\end{eqnarray}
A straightforward calculation yields:
\begin{align}
    B_{s}(0,|\vec{q}| = \lf, \lf) = \mathcal{A}^* \int_{\infty}^{\Lambda_f}{\rm d}\lambda \;  \lambda^{y} \frac{\Lambda_f}{\lambda^2} = \nonumber \\
    \mathcal{A}^* \frac{1}{y-1} \; \Lambda_f^{y} 
    \xrightarrow{\Lambda_f \to 0} 
    \begin{cases}
        0  \quad {\rm for} \; y\in (0, 1) \\
        const  \quad {\rm for} \; y = 0 \\
        -\infty  \quad {\rm for} \; y < 0 \\
    \end{cases} \; .
\end{align} 
As already noted, for $y>0$ the flow converges to a state with $Q_{\Lambda\to 0}>0$, while for $y<0$ we obtain an unphysical behavior of the mass. 
It follows that occurrence of a critical state (within the present truncation level) should require that 
\begin{eqnarray} \label{eq:semi:ycondition}
    y = - 2 \eta_g + \eta_f = 0 \; .
\end{eqnarray}
One may demonstrate that the present procedure then leads to the superficially appealing result 
\begin{equation}
    z_b = 3 x = 3(1 + \eta_f ) \;.
\end{equation}
However, inspection of the Fermi self-energy flow reveals a problem. 
We evaluate the flow of $Z_f$, and take note of the fact that, at the fixed point:  
\begin{equation}
    y = -2\eta_g + \eta_f = 0 \;\;\;\; \implies \;\;\;\; \frac{g_\Lambda^2}{Z_f} \to const \; ,
\end{equation}
which leads to 
\begin{align} \label{eq:freq:Zfsemi}
    \partial_{\Lambda_f} Z_f &\sim - 
    \frac{g_\Lambda^2 Z_f v_f^2 \Lambda_f}{2 \pi^3}
    \underbrace{\frac{ v_f \Lambda_f}{Z_f}}_{{\rm d}q_0}
    \underbrace{\frac{1}{v_f^4 \Lambda_f^4}}_{fermionic} \;
    \underbrace{\frac{1}{\Lambda_f^{2/3}}}_{bosonic} 
    \underbrace{\Lambda_f^{4/3}}_{{\rm d}q^2}
    \; \sim \notag \\
&-  \frac{g_\Lambda^2}{v_f \Lambda_f^{4/3} } \sim - \frac{A^* Z_f}{k_f \Lambda_f^{4/3}} \;\;.
\end{align}
This behavior is incompatible with the scaling assumed in Eq.~(\ref{parametrization}) for any finite value of $\eta_f$. 
%Notably, Fermi momentum appears above as a multiplication factor, which suggests
%the importance of renormalizing the momentum-dependent part of fermionic self-energy as well. We elaborate on this below in Sec.~IV.B.

In conclusion to this section: we proposed a step beyond the treatment discussed in Sec.~III, by self-consistently  introducing the scaling of frequency dependence of the self-energy as well as the Yukawa interaction coupling. At the same time we kept  the momentum dependencies in the Fermi self-energy  unrenormalized. The obtained result indicates no possibility of realizing a sensible non-Fermi liquid fixed point within this procedure.   
In what follows, we extend the truncation by additionally allowing for RG scaling of the Fermi velocity and demonstrate that this does not remove the inconsistency.  
%It is notable that the, for $y>0$ the ordering wavevector converges to a non-zero value.
\subsection{Scaling of $v_f$} \label{sec:vf}
\begin{figure}
    \centering
\includegraphics[width=\linewidth]{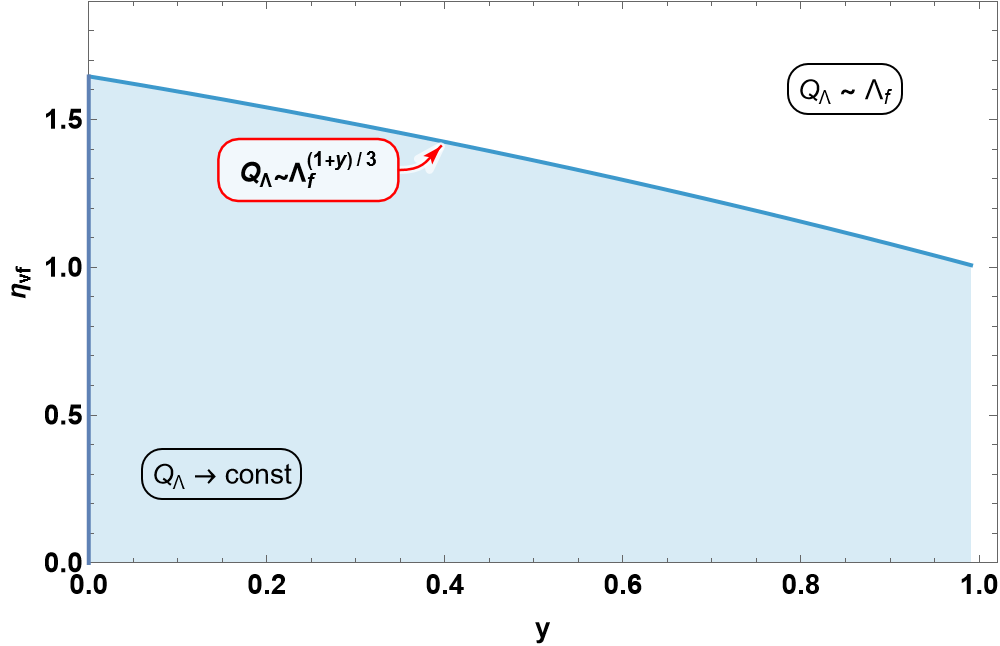}
    \caption{Schematic illustration of the behavior of $Q_\Lambda$ depending on $y$ and $\eta_{vf}$. 
    The solid blue line separating the two regimes follows from the condition $y\, b_0(y,\eta_{vf}) = 0$; see the main text for a discussion.}
    \label{fig:SCFL:L:asymQlarge}
\end{figure}

\begin{figure}
    \centering
\includegraphics[width=\linewidth]{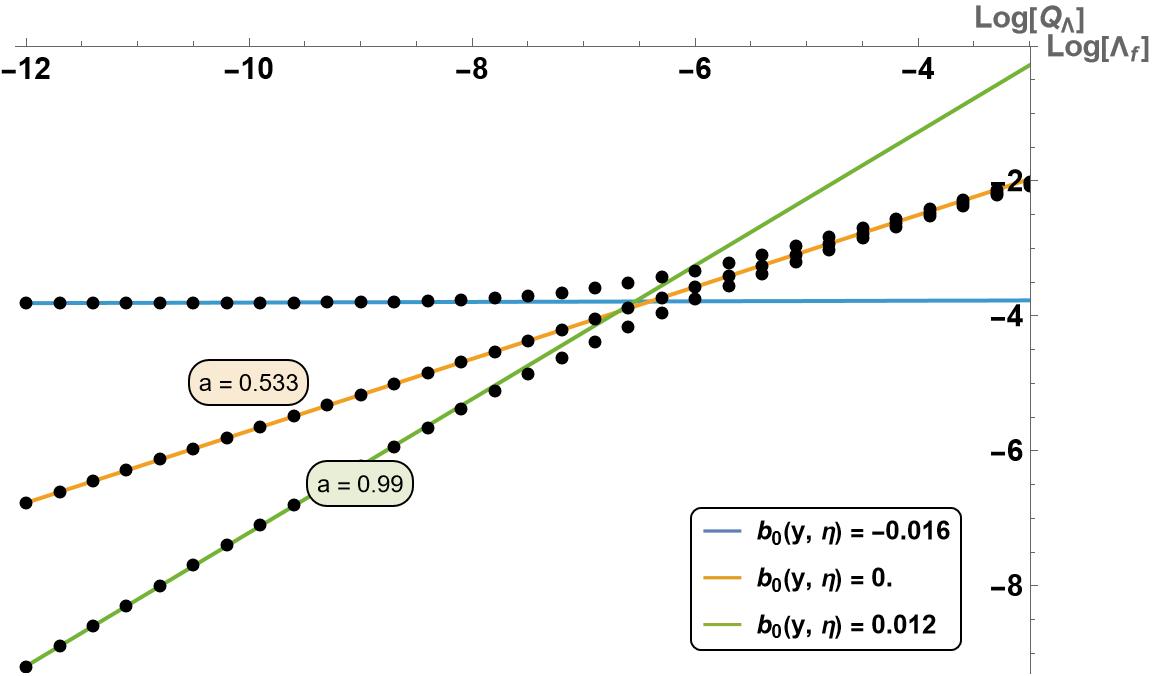}
    \caption{Exemplary illustration  of the scaling of the minimum $Q_\Lambda \sim \Lambda_f^a$ for different values of $b_0(y, \eta_{vf})$, keeping $y = 0.6$ fixed. The numerical results are plotted as black dots, while the lines represent fits in the region of small $\Lambda_f$. The three curves illustrate the three types of behavior described in the text: convergence to $Q_{\Lambda\to0}>0$, $Q_\Lambda\sim \Lambda_f^{(1+y)/3}$, and $Q_\Lambda\sim\Lambda_f$.}
    \label{fig:SCFL:N:minima}
\end{figure}

We now extend the truncation by additionally allowing  for the flow of the Fermi velocity. We  impose 
\begin{eqnarray}
        \xi_{k, \Lambda}= \dfrac{v_{f, \Lambda} }{2 k_f} (|\vec{k}|^2 - k_f^2) \ 
    \end{eqnarray}
and the fixed point scaling $v_{f, \Lambda} = v_f^* \; \Lambda_f^{-\eta_{vf}}$, which defines $\eta_{vf}$. We also introduce 
\begin{align}
    y := -2 \eta_g + \eta_f +\eta_{vf}\;, \;\;\;\;
    x := 1 + \eta_f - \eta_{vf} \; .
\end{align}
The flow of $v_{f, \Lambda}$ is reflected in the flowing bubble diagram via the redefinition of the regularized dispersion $f_{\Lambda}(\vec{k}) = \xi_{\vec{k}, \Lambda}+R_f(\vec{k})$ and via an entirely new term appearing in $\partial_{\Lambda} R_f$. The scaling form of the regularized bubble, here denoted by $\mathcal{X}_{sc}$, is given by:
\begin{eqnarray} \label{eq:SCFL:scalingCHI}
    \mathcal{X}_{sc}(q_0, q, \lf) &=& \nonumber \\ 
    \frac{\mathcal{A}^* \Lambda_f^{y}}{q} \;
    &\bigg[&
     (1 - \eta_{vf}) \,
    h\big( t, w \big) +
     \eta_{vf} \, h_{k^2}\big( t, w \big) 
    \bigg],
\end{eqnarray} 
where $t =\Lambda_f/q$ and $w = \tfrac{Z_{f}^*}{v_f^*} \tfrac{q_0}{q^x}$. The derivation of the above formula and the form of the function $h_{k^2}\big(t, w \big)$ are  presented in Appendix B (for a plot of $h_{k^2}(t)$ see also Fig.~\ref{fig:h_function}). 
Following the path described in the previous sections, we analyze the minimum of the inverse propagator and extract its   behavior for $\Lambda_f \to 0$. We obtain
\begin{eqnarray} \label{eq:SCFL:Qequation}
    &\dv{}{q}&\left( q^2 + B(0, q, \lf) \right)_{q = Q_\Lambda}
    =  \notag \\
    &=& 2 Q_\Lambda + \frac{y}{Q_{\Lambda}} B_{sc}(0, Q_{\Lambda}, \lf) - \nonumber \\
    &\mathcal{A}^*& \; \frac{\Lambda_f^{1+y}}{Q_\Lambda^2} \Big[
    (1 - \eta_{vf})\, h\Big(\frac{\Lambda_f}{Q_\Lambda}\Big) + \eta_{vf} \, h_{k^2}\Big(\frac{\Lambda_f}{Q_\Lambda}\Big)
    \Big]
    = 0 \; . \quad 
\end{eqnarray} 
We have performed a detailed analysis of the solutions to Eq.~(\ref{eq:SCFL:Qequation}) depending on $y$, $\eta_{vf}$, and $\Lambda_f$. Our analysis indicates a generic lack of possibility of obtaining an RG fixed point with properties reminiscent of the RPA-type behavior discussed in Sec.~III. Our conclusion is summarized in the diagram of Fig.~\ref{fig:SCFL:L:asymQlarge}. In the physically most interesting range of $y>0$ and $0<\eta_{vf}<1$ the flowing wavevector saturates for $\Lambda_f$ small at a positive value. In the other regime (where $\eta_{vf}>1$) we obtain a linear scaling of $Q_\Lambda$ as a function of $\Lambda_f$, which however is accompanied by appearance of another competing minimum of $G_b^{-1}$ directly at momentum $q=0$. This picture points at a possibility of a first-order transition and fragility of the system upon  further elevating  the truncation level. These two regimes are separated in  the diagram of Fig.~\ref{fig:SCFL:L:asymQlarge} by a line located at 
\begin{equation}
 y\, b_0(y,\eta_{vf}):=y  \int_{\infty}^{0} {\rm d} t \, t^y
    \big[ (1 - \eta_{vf}) h(t) + \eta_{vf} h_{k^2}(t)  \big]=0\;,
\end{equation}
where we find $Q_\Lambda\sim \Lambda_f^{(1+y)/3}$; in this case $G_b^{-1}$ also features two minima as function of $q$. The scaling of $Q_\Lambda$ is  
 exemplified in Fig.~\ref{fig:SCFL:N:minima} for the three cases discussed above. An entirely separate case occurs for $y=0$, where we identified a physical inconsistency in the flow of $Z_f$, analogous to the one discussed in Sec.~IV.A for $\eta_{vf}=0$. 
 
\begin{figure}
    \centering
\includegraphics[width=\linewidth]{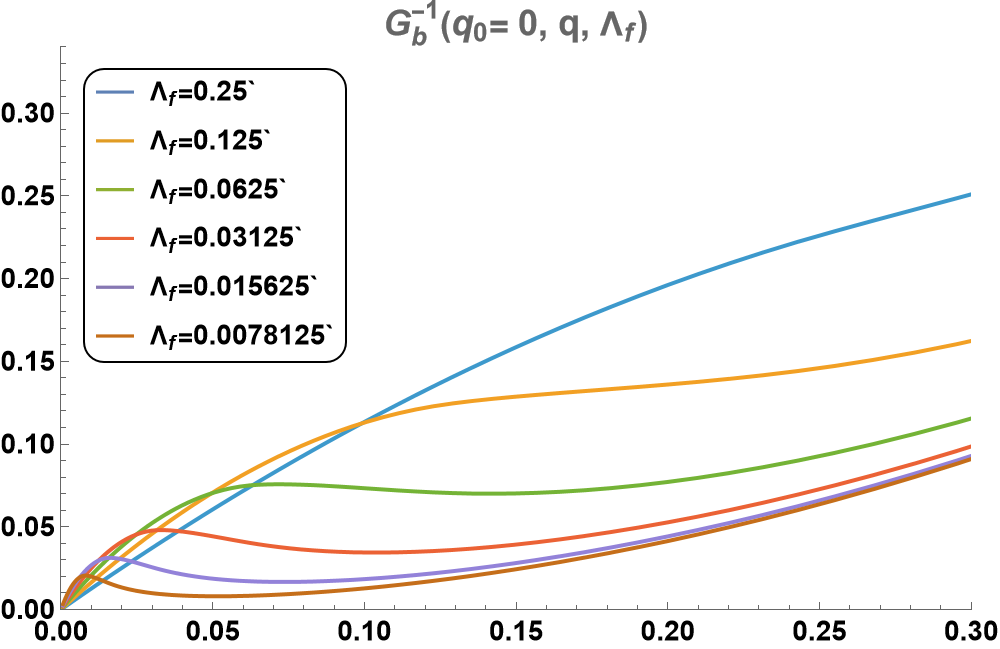}
    \caption{An exemplary plot of $G_b^{-1}(q,q_0=0,\Lambda_f)$ for a sequence of values of $\Lambda_f$ in a situation involving the $Q_\Lambda\sim \Lambda_f^{(1+y)/3}$ scaling. 
    The plot parameters are $y  = 0.6$ and $\eta_{vf} = 1.29568$, which leads to $b_0(y, \eta_{vf}) \approx 0$.}
    \label{fig:SCFL:N:inverseGb}
\end{figure}

To supplement the above numerical analysis we studied the asymptotic properties of Eq.~(\ref{eq:SCFL:Qequation}). Let us begin with the case  $Q_\l \ll \lf$. It is easy to show that Eq.~(\ref{eq:SCFL:Qequation})  features a solution given by 
\begin{eqnarray}
   Q_\Lambda = - \frac{4(1 - \eta_{vf}) (2 - y)}{\pi \,\eta_{vf} (1 - y)} \Lambda_f \; ,
\end{eqnarray}
the positivity of which imposes constraints on $\eta_{vf}$ and $y$. It turns out that this solution corresponds to a maximum of $G_b^{-1}$, which is visible in the curves plotted in Fig.~\ref{fig:SCFL:N:inverseGb}.

Another solution is identified by imposing $y \, b_0(y, \eta_{vf}) = 0$. In this case we find 
\begin{equation} \label{eq:SCFL:L:min}
    Q_\Lambda = \Big( \frac{3 \big(1 - \frac{\eta_{vf}}{2} \big) \mathcal{A}^*}{2 (1 + y)} \Big)^{\frac{1}{3}} \Lambda_f^{\frac{1 + y}{3}} \;, 
\end{equation} 
which fully agrees with the numerical data (see Figs.~\ref{fig:SCFL:L:asymQlarge} and \ref{fig:SCFL:N:minima}).
For $y = 0$ we retrieve the result of Sec.~IV.A, where $Q_\Lambda \sim \Lambda_f^{1/3}$. Finally, by expanding Eq.~(\ref{eq:SCFL:Qequation}) for the other scaling limit, $Q_\Lambda\gg \lf$, we obtain 
\begin{equation} 
\label{rob_2}
    Q_\Lambda^{y+1} \approx \frac{3 \big(1 - \frac{\eta_{vf}}{2} \big)}{y (y + 1) \, b_0(y, \eta_{vf})} \Lambda_f^{y+1} \;,
\end{equation}
leading to the scaling $Q_\Lambda\sim \Lambda_f$ as found in the numerical analysis. Note that the condition $Q_\Lambda\gg \lf$ remains fulfilled by Eq.~(\ref{rob_2}) due to the large value of the coefficient on the RHS. 

We summarize this section by the conclusion that consistent account of scaling of the momentum and frequency dependencies of the self-energy (parametrized by flowing $Z_f$ and $v_f$) as well as the Yukawa coupling $g$ does not allow for obtaining a sensible non-Fermi liquid fixed point with ordering wavevector $\vec{Q}=\vec{0}$. Instead, we identified a generic tendency towards renormalizing the ordering wavevector to a non-vanishing value or an inverse propagator featuring two minima (pointing towards possible first order transition) accompanied by a large value of the scaling exponent $\eta_{vf}$. 

Existence of a non-Fermi liquid fixed point within the self-consistent treatment is however not ruled out upon including higher order terms, which are absent in the initial action of Eq.~(\ref{bare_action}), but become generated by the RG flow. We speculate on this possibility below in Sec.~IV.C.

\subsection{Fermion-boson 4-point vertex} \label{sec:FBvertex}

%In systems with only bosonic degrees of freedom, such as the quantum Ising model, the contribution from the bosonic bubble diagram is compensated by a tadpole diagram consisting of a bosonic 4-point vertex and a single closed loop (see Fig.~\ref{fig:diagrams}). The analogous diagram for mixed fermion-boson systems is described by the following term:
Let us now consider the following contribution to the flow of the boson self-energy 
\begin{equation} \label{eq:ROUTE:tadpole}
    \mathcal{X}_{\gamma}(Q, \lf) = - 2 \int_K \ 
    S_f(K) \, \gamma_\Lambda(Q, K) \; ,
\end{equation}
which involves the 4-point fermion-boson vertex $\gamma_\Lambda$
%where function $\gamma_\Lambda$ defines the 4-point fermion-boson vertex:
\begin{equation}
    \gamma_\Lambda(Q, K) \coloneqq \dfrac{\delta^{4} \Gamma_{\Lambda}[\psi, \bar{\psi},\phi]}{ \delta \phi_{Q} \delta \phi_{-Q} \delta \psi_{K, \uparrow} \delta \bar{\psi}_{K, \uparrow}} \; .
\end{equation}
In standard treatments (including our own analysis thus far), this contribution is  neglected for several reasons. First, the 4-point vertex is absent in the bare action. Furthermore, evaluating it strictly at the minimum of the inverse propagator, $K = \{k_0 = 0, |\vec{k}| = k_f\}$, yields zero, since Eq.~(\ref{eq:ROUTE:tadpole}) involves an integral over a second-order pole. Therefore, any non-trivial result can only arise from the frequency-dependent part of $\gamma_\Lambda(Q, K)$ that is generated during the integration of the flow.

\begin{figure}
    \centering
    \includegraphics[width=0.32\linewidth]{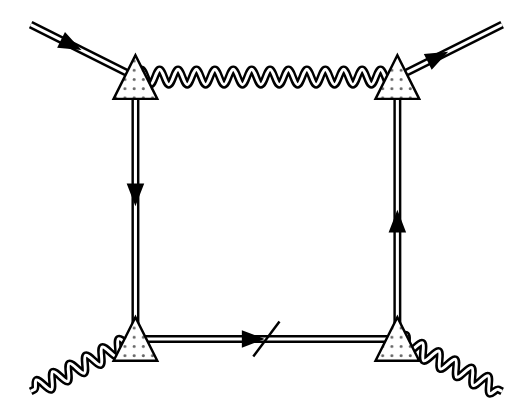}
    \includegraphics[width=0.32\linewidth]{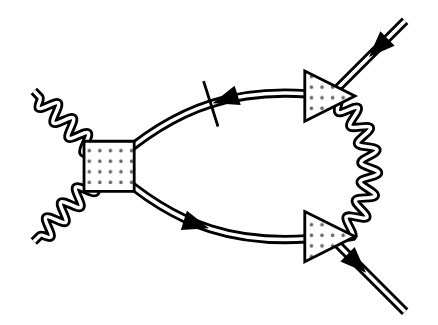}
    \includegraphics[width=0.32\linewidth]{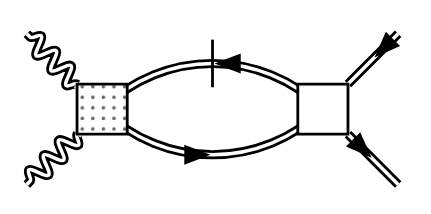}
    \caption{Exemplary diagrams which contribute to $\partial_\Lambda \gamma_\Lambda(Q,K)$.}
    \label{fig:g4diagrams}
\end{figure}
\begin{figure}
    \centering
    \includegraphics[width=1.\linewidth]{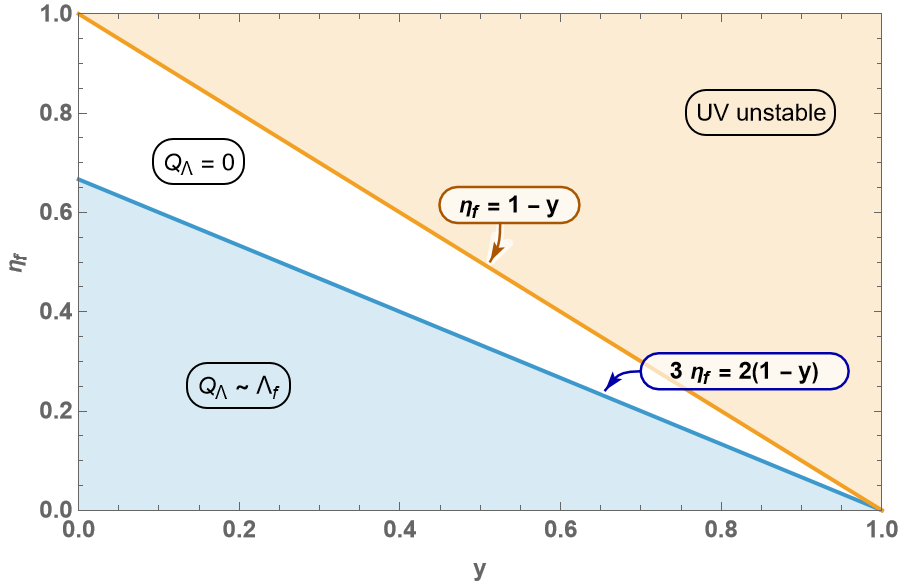}
    \caption{Scaling of the global minima $Q_\Lambda$ for different values of  $\eta_f$ and $y$. The most promising scaling from the point of view of restoring the QCP with $\vec{Q}=\vec{0}$ is marked by a light-blue region ($Q_\Lambda \sim \Lambda_f$) and dark blue line ($Q_\Lambda = 0$). The latter differs from the white region in momentum dependence of the bosonic propagator around minimum. The upper right corner of this plane (orange region) exhibits strong dependence on the UV cut-off. We do not explore this region in the present work.}
    \label{fig:ROUTE:summarized}
\end{figure}
\begin{figure}
    \centering
    \includegraphics[width=\linewidth]{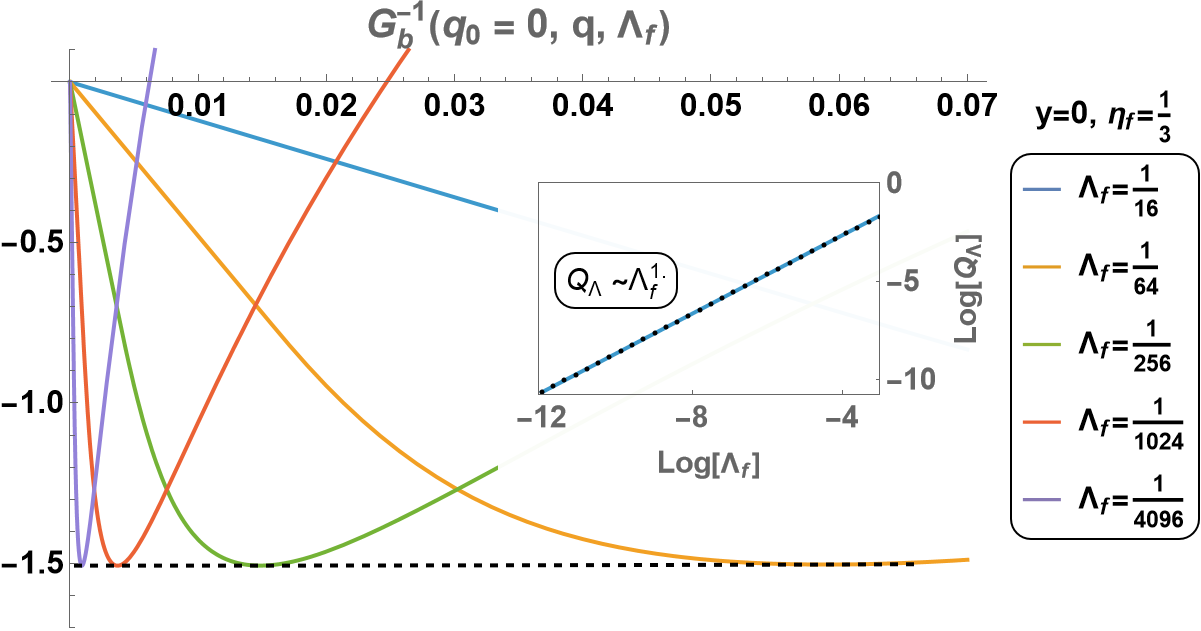}
    \includegraphics[width=\linewidth]{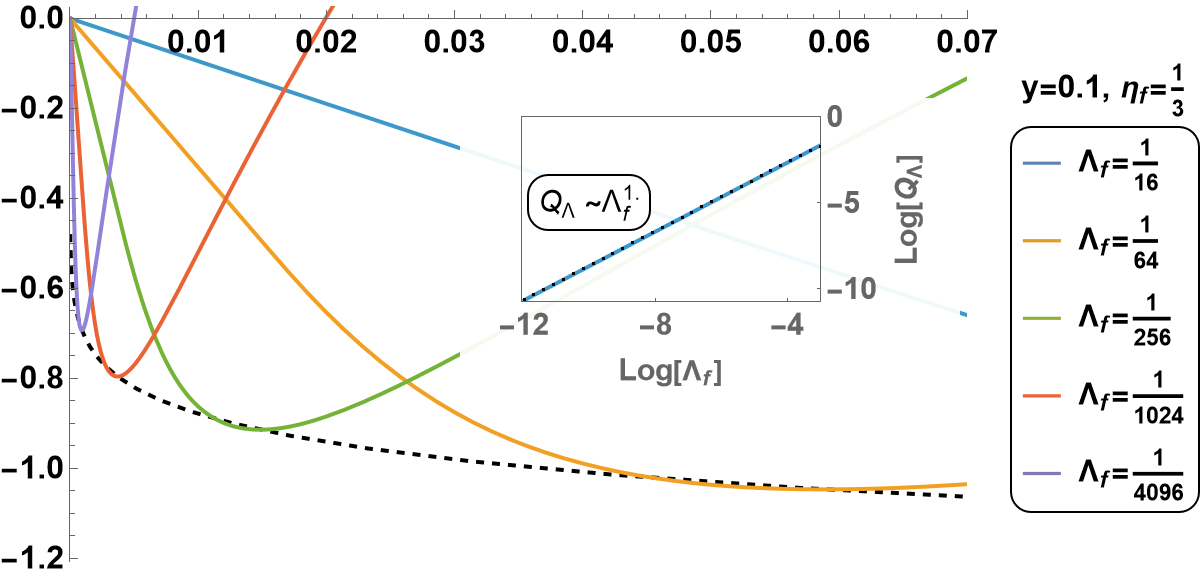}
    \caption{The  plots show  $G_b^{-1} (0,q, \lf)$ for different values of $\Lambda_f$. The black dashed line demonstrates   scaling of the  minimum of the inverse propagator with $\Lambda_f$. For $y>0$ we find a tendency towards $0$ (lower panel), while for $y= 0$ it flows to a constant value (upper panel). The log-log plots of $Q_\Lambda (\Lambda_f)$ are shown in the insets. }
    \label{fig:ROUTE:GinvLB}
\end{figure}

Fig.~\ref{fig:g4diagrams} depicts representative diagrams that contribute to $\partial_\Lambda \gamma_\Lambda(Q,K)$. Only the first of these --- the box diagram --- can be constructed from vertices present already in the bare action. Below we consider only this diagram. %neglecting all others, 
%and focus on the term involving $\mathcal{S}_f$. 
While we retain the scaling of the Yukawa coupling and the fermionic self-energy, for simplicity, we omit the flow of the Fermi velocity. Consequently, Eq.~(\ref{eq:ROUTE:tadpole}) can be rewritten as follows:
\begin{widetext}
\begin{equation} 
    \mathcal{X}_{\gamma}(Q, \lf) = 2 \int_{K_1} \ S_{f}(K_1, \lf)  
    \begin{multlined}[t]
    \int_{\Lambda_{f, UV}}^\lf {\rm d} \lambda 
    \Bigg\{
    g_{\lambda}^4 \int_{K_2} G_{f}^2(K_2, \lambda) G_{b}(K_1 - K_2, \lambda) \Big[ S_{f}(K_2 + Q, \lambda) + S_{f}(K_2 - Q, \lambda)
    \Big] + \\
    +
    2 g_{\lambda}^4 \int_{K_2} S_{f}(K_2, \lambda) G_{f}(K_2, \lambda) G_{b}(K_1 - K_2, \lambda) \Big[ G_{f}(K_2 + Q, \lambda) - G_{f}(K_2 - Q, \lambda) \Big]
    \Bigg\} \; .
    \end{multlined}
\end{equation}
\end{widetext}
By inserting the box diagram into the tadpole diagram contributing to the flow of the  boson self-energy, we introduced two distinct flowing scales, $\lf$ and $\lambda$. We distinguish the scale dependence of the propagators explicitly. As in the previous section, we take the limit $\Lambda_{f,UV} \to \infty$. Collecting all terms that depend on the integration variable $K_1$, we define:
\begin{equation}
    W(K, \lambda, \lf) \coloneqq - g_{\lambda}^2 \int_{K_1} S_{f}(K_1, \lf) G_{b}(K_1 - K, \lambda) \; .
\end{equation}
Notice that for $\lambda = \lf$, this function resembles the flow of the fermionic self-energy [see Eq.~(\ref{eq:sigma})]. We can therefore approximate this term as:
\begin{equation}
    W(K, \lambda, \lf) \approx  - i \partial_{\lf} Z_{f, \Lambda_f} k_0 = i \, \eta_f \frac{Z_{f, \Lambda_f}}{\Lambda_f Z_{f, \lambda}} Z_{f, \lambda} k_0 \; .
\end{equation}
Substituting this result and taking the limit $k_f \gg max(\Lambda,\lf, q)$ we obtain the static expression:
\begin{align} \label{eq:xigamma}
    \mathcal{X}_{\gamma}(0, q, \lf) &\approx 2 \eta_f \frac{Z_{f, \Lambda_f}}{\Lambda_f}
    \int_{\infty}^{\Lambda_f} {\rm d} \lambda
    \frac{\mathcal{A_\lambda}}{Z_{f, \lambda}q}
    f\Big(\frac{\lambda}{q}\Big) \;.
\end{align}
For the definition and details of the scaling function $f(t)$, see Appendix F. The contribution from this vertex to the bosonic propagator is obtained by integrating over the scale:
\begin{flalign}
    B_\gamma(0, q, \lf)  &= \int_{+\infty}^{\Lambda_f}{\rm d} \lambda \mathcal{X}_{\gamma}(q_0 = 0, q, \lambda) \approx && \\
    &\approx 2 \mathcal{A}^* q^y \int_{+\infty}^{\Lambda_f/q}{\rm d} t \,
    t^y f(t)
    \left(1 
    - \Big(\frac{t q}{\Lambda_f}\Big)^{\eta_f}
    \right) \; . &&
\end{flalign}
As before, we determine the flowing ordering wavevector by solving for the extremum of the bosonic propagator:
\begin{align} 
\label{ght}
    \dv{ }{q} \left(  q^2 + B_s(0,q, \lf) + B_{\gamma}(0, q, \lf)\right)_{q = Q_\Lambda} = 0 \; .
\end{align} 
Inspection of the structure of $B_{\gamma}(0, q, \lf)$ reveals that it tends to reduce the magnitude of $Q_\Lambda$ towards zero.
The numerical results of solution of Eq.~(\ref{ght}) are summarized in Fig.~\ref{fig:ROUTE:summarized}. For small values of the exponent, $\eta_f < 2/3 (1 - y)$, we find a linear dependence of $Q_\Lambda$ on  the fermionic cutoff scale $\Lambda_f$:
\begin{align}
    Q_\Lambda = C_1 \Lambda_f \qquad {\rm where} \quad C_1 > 2 \; .
\end{align}
At the boundary of the blue region (for $\eta_f = 2/3 (1 - y)$), the global minimum is located at $q = 0$. For $\eta_f \geq 1 - y$, the procedure of sending $\Lambda_{f, UV}$ to infinity is no longer valid.
Starting from the RPA calculation, we focus on the region where $y$ is small and $\eta_f \approx 1/3$. We observe that the predicted scaling of $Q_\Lambda$ departs from the previously obtained result, $Q_\Lambda \sim \Lambda_f^{1/3}$, thereby avoiding the problems encountered in the flow of the fermionic self-energy presented in Eq.~(\ref{eq:freq:Zfsemi}). Within the present analysis, the value of $z_b$ remains undetermined, as it depends heavily on the precise values of $y$ and $\eta_f$. The role of the omitted contributions to $\gamma_\Lambda(Q,K)$ is also out of present control and is relegated to future studies. 
In Fig.~\ref{fig:ROUTE:GinvLB}, we depict the full inverse bosonic propagator $G_b^{-1}$ in the static limit, accounting for mass renormalization.

 As demonstrated in Sec.~III, flow of $Q_\Lambda$ appears essential for a correct account of Landau damping within Wilsonian-type RG. On the other hand it gives rise to the instability of the non-Fermi liquid fixed point in the improved truncations discussed in Sec.~IV.A and B, which is reflected in the flow of $Q_\Lambda$ towards positive values. This may however, perhaps, change when also quartic interaction vertices (including their frequency dependencies) are considered, as signaled in the present section.   

\section{Summary} 
Quantum criticality of two-dimensional Fermi systems has since long remained a topic of top relevance and interest  to condensed matter theory. Despite substantial efforts over many years, a fully satisfactory theory  of these phenomena has not emerged up to this point. In the present paper we have proposed a consistent approach to this problem from the  Wilsonian RG point of view, adapting the Wetterich equation as the point of departure. In the first place, remaining at the functional RG level, we have shown how a relatively simple, non-self-consistent truncation of the Wetterich equation allows for obtaining a non-Fermi liquid  RG fixed point with 
the features anticipated from earlier one-loop-level studies. We also tested the validity of simplified treatments, which abandon the computation of the full frequency dependence of the self-energy in favor of a commonly implemented simplified parametrization capturing only the low energy asymptotics.
Our approach regularizes all the degrees of freedom present in the theory and evades the problem of singular effective bosonic interactions. In particular, the boson propagator in the loop integrals is only dressed by fluctuations above the cutoff scale, leading to a structure entirely different as compared to the singular fully dressed Hertz-Millis form commonly used in many earlier approaches. An unavoidable feature emerging in our scheme is the RG flow of the ordering wavevector $\vec{Q}_\Lambda$, a property which cures  problems of the mass flow encountered in earlier Wilsonian RG studies. 
The fixed-point scaling of $Q_\Lambda$ determines the value of the bosonic dynamical exponent $z_b=3$ and the Fermi self-energy exponent $\alpha=2/3$. 
Previous perturbative approaches correct the value of $\alpha$ only at the 3-loop level and yield unusual and very unclear predictions concerning $z_b$ at 4 loops. 

In the second part of our paper (Sec.~IV) we set out to investigate  how inclusion of nontrivial scaling of the self-energy and the Yukawa coupling impact the RG flow. We have demonstrated an instability of the non-Fermi liquid RG fixed point towards either a phase featuring $\vec{Q}\neq \vec{0}$ or a behavior typical to first-order transitions. The latter is 
expected to generically occur in the case of magnetic transitions \cite{Belitz_2005, Kirkpatrick_2026}, and, at least at mean-field level, is also most common for electronic nematics \cite{Yamase_2005}. Despite these findings, occurrence of a QCP featuring $\vec{Q}=\vec{0}$ is not excluded, but requires accounting for higher-order terms and would presumably imply critical singularities quite distinct from those characteristic to one-loop level. In Sec.~IV.C we presented a plausible, but, at the present point, speculative picture of how critical scaling might reemerge by including terms involving boson-fermion interaction vertices, which are entirely  absent in the bare action, but become generated by the RG flow.  

The role of the flow of $Q_\Lambda$, which was  disregarded in previous Wilsonian RG approaches to this problem, is striking. On one hand, it appears as an indispensable part of the procedure and its behavior for $\Lambda\to 0$ determines the scaling exponents at the leading order of the calculation (Sec.~III). On the other hand, within the most natural, lowest-order self-consistent extensions of the truncation (Sec.~IV) its generic flow to a not-zero value gives an indication of an instability of the RPA-type non-Fermi liquid state. It remains open whether the instabilities obtained in Sec.~IV.A and B represent the true physical picture (in agreement with previous literature claims \cite{Chubukov_2004, Belitz_2005,  Kirkpatrick_2026} concerning the generic instability of $\vec{Q}=\vec{0}$ QCPs), or the RG fixed point becomes restored at a significantly higher level of truncation (for example by the mechanism presented in Sec.~IV.C). In the former case we expect a fixed point signinficantly different from the one obtained at RPA level.          

\begin{acknowledgments}
 We are grateful to Walter Metzner for very useful discussions and to Thomas Sheerin for  remarks on the manuscript. M.H. and P.J. acknowledge support from the Polish National Science Center via grant 2021/43/B/ST3/01223. H.Y. was supported by JSPS KAKENHI Grant No.JP20H01856 and
World Premier International  Research Center Initiative (WPI), MEXT, Japan.
\end{acknowledgments} 

\newpage

\bibliography{bibliography}% Produces the bibliography via BibTeX.

\clearpage

\appendix

\begin{widetext}

\setcounter{figure}{0}                       % <---------------
\renewcommand\thefigure{B.\arabic{figure}}   % <---------------

\section{Derivation of the Wetterich equation}

We summarize the derivation of the Wetterich equation for a mixed fermion-boson system folowing  Ref.~\cite{Obert_2014_PhD}. An alternative but fully equivalent supertrace formalism can be found in Refs.~\cite{Kopietz_book, Metzner_2012}.

We consider a general form of the bare action:
\begin{equation}
S[\tilde{\psi}, \tilde{\phi}] = 
\frac{1}{2} \int_{\K, \K'} \tilde{\psi}_{\K} \mathcal{G}^{-1}_f(\K, \K') \tilde{\psi}_{\K'}
+ \frac{1}{2} \int_{\Q, \Q'} \tilde{\phi}_{\Q} \mathcal{G}^{-1}_b(\Q, \Q') \tilde{\phi}_{\Q'} +
V[\tilde{\psi}, \tilde{\phi}] \;,
\end{equation}
where we have extended the notation for the fermionic and bosonic variables to collect frequencies ($k_0$ or $q_0$), momenta ($\vec{k}$ or $\vec{q}$), the spin index $\sigma$, and the Grassmann conjugation index $c$ for fermions (or complex conjugation for bosons), such that $\K = \{ K, \sigma, c\}$ and $\Q = \{ Q, c \}$. In this notation, the fermionic field is represented by the vector $\psi_{\K} = [ \psi_{K, \uparrow} , \psi_{K,  \downarrow} , \bar{\psi}_{K, \uparrow} , \bar{\psi}_{K, \downarrow} ]$, while the bosonic field is given by $\phi_\Q = [\phi_Q \, \phi_{-Q} ]$, since the order parameter in the present work is represented by a real field.  
% Note that the anticommutation and commutation relations for the fermionic and bosonic fields, respectively, impose the following symmetry properties on the propagators:
% \begin{align}
%     \mathcal{G}^{-1}_f(\K, \K') = - \mathcal{G}^{-1}_f(\K', \K) & &
%     \mathcal{G}^{-1}_b(\Q, \Q') = \mathcal{G}^{-1}_b(\Q', \Q) \; .
% \end{align}
We introduce scale-dependent cutoff functions parameterized by $\l$:
\begin{align}
    \Delta S_\l [\tilde{\psi}, \tilde{\phi}] = 
    \frac{1}{2} \int_{\K, \K'} \tilde{\psi}_{\K} \mathcal{R}^{-1}_f(\K, \K') \tilde{\psi}_{\K'}
+ \frac{1}{2} \int_{\Q, \Q'} \tilde{\phi}_{\Q} \mathcal{R}^{-1}_b(\Q, \Q') \tilde{\phi}_{\Q'} \; .
\end{align}
Adding these regulator terms along with source terms for the external fields $J_\Q$ and $\eta_\K$, we define the scale-dependent partition function:
\begin{equation}
    Z_\l[\eta, J] \coloneq \int \mathcal{D}\tilde{\psi} \mathcal{D}\tilde{\phi} 
    \exp \left( -S[\tilde{\psi}, \tilde{\phi}]  - \Delta S_\l [\tilde{\psi}, \tilde{\phi}] + \int_\Q J_\Q \tilde{\phi}_\Q  + 
    \int_K \eta_\K \tilde{\psi}_\K \right) \; .
\end{equation}
The logarithm of the partition function 
\begin{equation}
    W_\l[\eta, J] = \ln \left( Z_\l[\eta, J] \right)
\end{equation}
serves as the generating functional for connected correlation functions, the expectation values of the fields are obtained via:
\begin{align} \label{app1:fields}
    \phi_\Q = \langle \tilde{\phi}_\Q \rangle = \frac{\delta W_\l[\eta, J]}{\delta J_\Q} & & \psi_\K = \langle \tilde{\psi}_\K \rangle = \frac{\delta W_\l[\eta, J]}{\delta \eta_\K} \; .
\end{align}
We perform a Legendre transform of $W_\l$ to define the effective action:
\begin{align}
    \tilde{\Gamma}_\l[\psi, \phi] \coloneq - W_\l[\psi, \phi] + \int_\Q J_\Q \phi_Q + \int_\K \eta_\K \psi_\K \; .
\end{align}
This new functional serves as the generating functional for one-particle-irreducible (1PI) vertex functions, such that:
\begin{align} \label{app1:fields2}
    \eta_\K = - \frac{\delta \tilde{\Gamma}_\l[\psi, \phi]}{\delta \psi_\K} &&
    J_\Q = \frac{\delta \tilde{\Gamma}_\l[\psi, \phi]}{\delta \phi_\Q} \; .
\end{align}
We can straightforwardly derive the flow equation for the effective action by taking the total derivative with respect to the scale $\l$:
\begin{align}
    \dv{}{\l} \tilde{\Gamma}_\l[\psi, \phi] = 
    - \partial_\l W_\l[\eta, J] = \frac{1}{2} \int_{\Q, \Q'} \partial_\l \mathcal{R}_b(\Q, \Q') \langle \tilde{\phi}_\Q \tilde{\phi}_{\Q'} \rangle + \frac{1}{2} \int_{\K, \K'} \partial_\l \mathcal{R}_f(\K, \K') \langle \tilde{\psi}_\K \tilde{\psi}_{\K'} \rangle  \; .
\end{align}
Modifying the Legendre transform by subtracting the regulator term, $\Gamma_\l[\psi, \phi] \coloneq \tilde{\Gamma}_\l[\psi, \phi] - \Delta S[\psi, \phi]$, allows us to rewrite the flow equation in the form:
% \begin{align}
%     \dv{}{\l} \Gamma_\l[\psi, \phi] &= \frac{1}{2} \int_{\Q, \Q'} \partial_\l \mathcal{R}_b(\Q, \Q') \frac{\delta^2 W_\l[\eta, J]}{\delta J_\Q \delta J_{\Q'}} + 
%     \frac{1}{2} \int_{\K, \K'} \partial_\l \mathcal{R}_f(\K, \K') \frac{\delta^2 W_\l[\eta, J]}{\delta \eta_\K \delta \eta_{\K'}}  \\[1.5ex]
%     &= \frac{1}{2} {\rm Tr} \left[ (\partial_\l \mathcal{R}_b) \frac{\delta^2 W_\l[\eta, J]}{\delta J^2} \right] - 
%     \frac{1}{2} {\rm Tr} \left[ (\partial_\l \mathcal{R}_f) \frac{\delta^2 W_\l[\eta, J]}{\delta \eta^2} \right] \; .
% \end{align}
% In the last line, we have used matrix notation for brevity. As a final step, we must express the second derivatives of $W_\l$ in terms of the derivatives of the effective action $\tilde{\Gamma}_\l$. This is achieved by differentiating Eq.~(\ref{app1:fields}) with respect to the fields $\psi$ and $\phi$, and substituting the relations from Eq.~(\ref{app1:fields2}). As a result, we obtain the exact flow equation:

% \begin{align} \label{app1:Wetterich}
%     \dv{}{\l} \Gamma_\l[\psi, \phi] &= 
%     \frac{1}{2} {\rm Tr} \left[ \big( \partial_\l \mathcal{R}_b \big) \Big(\frac{\delta^2 \tilde{\Gamma}_\l}{\delta \phi^2} \Big)^{-1} \Big( 1 - E_1 \Big)^{-1}
%     \right] +
%     \frac{1}{2} {\rm Tr} \left[ \big( \partial_\l \mathcal{R}_f \big) \Big(\frac{\delta^2 \tilde{\Gamma}_\l}{\delta \psi^2} \Big)^{-1} \Big( 1 - E_2 \Big)^{-1}
%     \right] \; .
% \end{align}

\begin{align}
    \dv{}{\l} \Gamma_\l[\psi, \phi] &= \frac{1}{2} \int_{\Q, \Q'} \partial_\l \mathcal{R}_b(\Q, \Q') \frac{\delta^2 W_\l[\eta, J]}{\delta J_\Q \delta J_{\Q'}} + 
    \frac{1}{2} \int_{\K, \K'} \partial_\l \mathcal{R}_f(\K, \K') \frac{\delta^2 W_\l[\eta, J]}{\delta \eta_\K \delta \eta_{\K'}}   \; .
\end{align}
As a final step, we express the second derivatives of $W_\l$ in terms of the derivatives of the effective action $\tilde{\Gamma}_\l$. This is achieved by differentiating Eq.~(\ref{app1:fields}) with respect to the fields $\psi$ and $\phi$, and substituting the relations from Eq.~(\ref{app1:fields2}). As a result, we obtain the exact flow equation:

\begin{align} \label{app1:Wetterich}
    \dv{}{\l} \Gamma_\l[\psi, \phi] &= 
    \frac{1}{2} {\rm Tr} \left[ \big( \partial_\l \mathcal{R}_b \big) 
    \Big( \tilde{\Gamma}^{(2)}_{\phi \phi} \Big)^{-1}
    \Big( 1 - E_1 \Big)^{-1}
    \right] +
    \frac{1}{2} {\rm Tr} \left[ \big( \partial_\l \mathcal{R}_f \big) 
    \Big( \tilde{\Gamma}^{(2)}_{\psi \psi} \Big)^{-1}
    \Big( 1 - E_2 \Big)^{-1}
    \right] \; .
\end{align}
Here we used a compat matrix notation, 
$ \Gamma^{(n)}_{\alpha_1 ... \alpha_n} \coloneqq \dfrac{\delta^{n} \Gamma_{\l}[\psi,\phi]}{ \delta \alpha_{1} ... \delta \alpha_{n}} $.
The  mixing between the bosonic and fermionic sectors is contained in the quantities $E_1$ and $E_2$:
% \begin{align}
%     & E_1 = \frac{\delta^2 \tilde{\Gamma}_\l}{ \delta \phi \delta \psi}  \Big(\frac{\delta^2 \tilde{\Gamma}_\l}{\delta \psi^2} \Big)^{-1}
%     \frac{\delta^2 \tilde{\Gamma}_\l}{ \delta \psi \delta \phi}
%     \Big(\frac{\delta^2 \tilde{\Gamma}_\l}{\delta \phi^2} \Big)^{-1} 
%     & E_2 = \frac{\delta^2 \tilde{\Gamma}_\l}{ \delta \psi \delta \phi}  \Big(\frac{\delta^2 \tilde{\Gamma}_\l}{\delta \phi^2} \Big)^{-1}
%     \frac{\delta^2 \tilde{\Gamma}_\l}{ \delta \phi \delta \psi}
%     \Big(\frac{\delta^2 \tilde{\Gamma}_\l}{\delta \psi^2} \Big)^{-1} \; .
% \end{align}
\begin{align}
    & E_1 = 
    \tilde{\Gamma}^{(2)}_{\phi \psi}
    \Big( \tilde{\Gamma}^{(2)}_{\psi \psi} \Big)^{-1}
    \tilde{\Gamma}^{(2)}_{\psi \phi}
    \Big( \tilde{\Gamma}^{(2)}_{\phi \phi} \Big)^{-1}
    & E_2 = 
    \tilde{\Gamma}^{(2)}_{\psi \phi}
    \Big( \tilde{\Gamma}^{(2)}_{\phi \phi} \Big)^{-1}
    \tilde{\Gamma}^{(2)}_{\phi \psi}
    \Big( \tilde{\Gamma}^{(2)}_{\psi \psi} \Big)^{-1} \; .
\end{align}
% Note that $E_1$ and $E_2$ contain vertices with an odd number of Grassmann fields; thus, they vanish when evaluated at a constant background field. However, these terms are necessary to generate the flow of higher-order vertices.
The flow equation for the effective action, presented in Eq.~(\ref{app1:Wetterich}) as a generating functional for 1PI vertices, was first derived by Wetterich \cite{Wetterich_1993} for a purely bosonic system. The flow of specific vertices can be obtained by taking appropriate functional derivatives with respect to the fields $\psi$ or $\phi$. By construction, the resulting flow equations contain only a single internal momentum integration and can therefore be represented by one-loop diagrams. All information regarding the flow is encoded in the structure of the fully dressed, scale-dependent propagators and vertices. %Compared to the supertrace notation, the formalism presented here explicitly separates the fermionic and bosonic single-scale operators. 
Finally, we provide the flow equations for self-energies:

\begin{align} \label{app1:Gamma2bos}
    \dv{}{\l} \Gamma^{(2)}_{\phi_{Q} \phi_{-Q}} =
    \pdv{\Pi(Q)}{\l} =&
    \frac{1}{2} {\rm Tr} \left[ \big( \partial_\l \mathcal{R}_b \big) 
    \Big( \tilde{\Gamma}^{(2)}_{\phi \phi} \Big)^{-1} 
    \Big( \tilde{\Gamma}^{(3)}_{\phi_{Q}  \phi \phi} \Big)
    \Big( \tilde{\Gamma}^{(2)}_{\phi \phi} \Big)^{-1}
    \Big( \tilde{\Gamma}^{(3)}_{\phi_{-Q}  \phi \phi} \Big)
    \Big( \tilde{\Gamma}^{(2)}_{\phi \phi} \Big)^{-1} 
    + (Q \to -Q)
    \right] + \nonumber \\
    & - \frac{1}{2} {\rm Tr} \left[ \big( \partial_\l \mathcal{R}_b \big) 
    \Big( \tilde{\Gamma}^{(2)}_{\phi \phi} \Big)^{-1} 
    \Big( \tilde{\Gamma}^{(4)}_{\phi_{Q} \phi_{-Q}  \phi \phi} \Big)
    \Big( \tilde{\Gamma}^{(2)}_{\phi \phi} \Big)^{-1}
    \right] + \nonumber \\    
    & + \frac{1}{2} {\rm Tr} \left[ \big( \partial_\l \mathcal{R}_f \big)
    \Big( \tilde{\Gamma}^{(2)}_{\psi \psi} \Big)^{-1}
    \Big( \tilde{\Gamma}^{(3)}_{\phi_{Q} \psi  \psi} \Big)
    \Big( \tilde{\Gamma}^{(2)}_{\psi \psi} \Big)^{-1}
    \Big( \tilde{\Gamma}^{(3)}_{\phi_{-Q} \psi \psi} \Big)
    \Big( \tilde{\Gamma}^{(2)}_{\psi \psi} \Big)^{-1}
    +(Q \to -Q)
    \right] \nonumber \\
    & - \frac{1}{2} {\rm Tr} \left[ \big( \partial_\l \mathcal{R}_f \big)
    \Big( \tilde{\Gamma}^{(2)}_{\psi \psi} \Big)^{-1}
    \Big( \tilde{\Gamma}^{(4)}_{\phi_{Q} \phi_{-Q} \psi \psi} \Big)
    \Big( \tilde{\Gamma}^{(2)}_{\psi \psi} \Big)^{-1}
    \right] 
\end{align}
and
\begin{align} \label{app1:Gamma2fer}
    \dv{}{\l} \Gamma^{(2)}_{\psi_{K, \uparrow} \bar{\psi}_{K, \uparrow}} =
    \pdv{\Sigma(K)}{\l} =&
    \frac{1}{2} {\rm Tr} \left[ \big( \partial_\l \mathcal{R}_b \big) 
    \Big( \tilde{\Gamma}^{(2)}_{\phi \phi} \Big)^{-1} 
    \Big( \tilde{\Gamma}^{(3)}_{\bar{\psi}_{K, \uparrow} \phi \psi} \Big)
    \Big( \tilde{\Gamma}^{(2)}_{\psi \psi} \Big)^{-1}
    \Big( \tilde{\Gamma}^{(3)}_{\psi_{K, \uparrow} \psi \phi} \Big)
    \Big( \tilde{\Gamma}^{(2)}_{\phi \phi} \Big)^{-1} 
    - (\psi_{K, \uparrow} \leftrightarrow \bar{\psi}_{K, \uparrow})
    \right] + \nonumber \\
    & - \frac{1}{2} {\rm Tr} \left[ \big( \partial_\l \mathcal{R}_b \big) 
    \Big( \tilde{\Gamma}^{(2)}_{\phi \phi} \Big)^{-1} 
    \Big( \tilde{\Gamma}^{(4)}_{\psi_{K, \uparrow} \bar{\psi}_{K, \uparrow} \phi \phi} \Big)
    \Big( \tilde{\Gamma}^{(2)}_{\phi \phi} \Big)^{-1}
    \right] + \nonumber\\    
    & + \frac{1}{2} {\rm Tr} \left[ \big( \partial_\l \mathcal{R}_f \big)
    \Big( \tilde{\Gamma}^{(2)}_{\psi \psi} \Big)^{-1}
    \Big( \tilde{\Gamma}^{(3)}_{\bar{\psi}_{K,\uparrow},\psi, \phi} \Big)
    \Big( \tilde{\Gamma}^{(2)}_{\phi \phi} \Big)^{-1}
    \Big( \tilde{\Gamma}^{(3)}_{\psi_{K, \uparrow} \phi \psi} \Big)
    \Big( \tilde{\Gamma}^{(2)}_{\psi \psi} \Big)^{-1}
    - (\psi_{K, \uparrow} \leftrightarrow \bar{\psi}_{K, \uparrow})
    \right] \nonumber \\
    & - \frac{1}{2} {\rm Tr} \left[ \big( \partial_\l \mathcal{R}_f \big)
    \Big( \tilde{\Gamma}^{(2)}_{\psi \psi} \Big)^{-1}
    \Big( \tilde{\Gamma}^{(4)}_{\psi_{K, \uparrow} \phi_{K, \uparrow} \psi \psi} \Big)
    \Big( \tilde{\Gamma}^{(2)}_{\psi \psi} \Big)^{-1}
    \right] \; .
\end{align}
The above equations are general and  exact. In the implementation to the present problem, we specify the
 explicit forms of the fermionic matrices in given basis as:
\begin{align}
\mathcal{R}_f
= R_f(K_1) \, \delta(K_1 - K_2)
    \begin{bmatrix}
    0 & 0 & - 1 & 0 \\
    0 & 0 & 0 & - 1 \\
    1 & 0 & 0 & 0 \\
    0 & 1 & 0 & 0
    \end{bmatrix}
 && 
\tilde{\Gamma}^{(2)}_{\psi \psi}
= - G_f^{-1}(K_1) \, \delta(K_1 - K_2)
    \begin{bmatrix}
    0 & 0 & - 1 & 0 \\
    0 & 0 & 0 & - 1 \\
    1 & 0 & 0 & 0 \\
    0 & 1 & 0 & 0
    \end{bmatrix}
\end{align}

\begin{align}
    \tilde{\Gamma}^{(3)}_{\phi_{Q} \psi  \psi} =
        \begin{bmatrix}
    0 & 0 & \tilde{\Gamma}^{(3)}_{\phi_{Q} \psi_{K_2, \uparrow} \bar{\psi}_{K_1, \uparrow}} & 0 \\
    0 & 0 & 0 & \tilde{\Gamma}^{(3)}_{\phi_{Q} \psi_{K_2, \downarrow} \bar{\psi}_{K_1, \downarrow}} \\
    -\tilde{\Gamma}^{(3)}_{\phi_{Q} \psi_{K_1, \uparrow}  \bar{\psi}_{K_2, \uparrow}} & 0 & 0 & 0 \\
    0 & -\tilde{\Gamma}^{(3)}_{\phi_{Q} \psi_{K_1, \downarrow} \bar{\psi}_{K_2, \downarrow}} & 0 & 0
    \end{bmatrix}
\end{align}

\begin{align}
    \tilde{\Gamma}^{(4)}_{\phi_{Q} \phi_{-Q} \psi \psi} = \tilde{\Gamma}^{(4)}_{\phi_{Q} \phi_{-Q} \psi_{K_2, \uparrow} \bar{\psi}_{K_1, \uparrow}} \delta(K_1 - K_2) 
        \begin{bmatrix}
    0 & 0 & 1 & 0 \\
    0 & 0 & 0 & 1 \\
    -1 & 0 & 0 & 0 \\
    0 & -1 & 0 & 0
    \end{bmatrix} &&
    \tilde{\Gamma}^{(3)}_{\psi_{K, \uparrow}  \phi\psi} =
    \begin{bmatrix}
    \tilde{\Gamma}^{(3)}_{\phi_{Q} \psi_{K, \uparrow} \bar{\psi}_{K_1, \uparrow}} & 0 & 0 & 0 \\
    \tilde{\Gamma}^{(3)}_{\phi_{-Q} \psi_{K, \uparrow} \bar{\psi}_{K_1, \uparrow}} & 0 & 0 & 0 \\
    \end{bmatrix} \; .
\end{align}

\section{The flowing bubble diagram and the scaling functions $h(t, w)$ and $h_{k^2}(t, w)$ }

This appendix provides the definitions, key derivation steps, and final forms of the scaling functions $h(t, w)$ and $h_{k^2}(t, w)$. These functions govern the flow of the fermionic bubble diagram $\mathcal{X}(Q, \Lambda_f)$ in the limit where the Fermi momentum is large compared to all other momentum scales.

The derivation presented below applies to the most general approximation discussed in this paper --- $\mathcal{X}_{sc}(Q, \Lambda_f)$ in section IV.B. The function $\mathcal{X}_s(Q, \lf)$ used in section IV.A. and IV.C can be retrieved by replacing the flowing velocity $v_{f, \Lambda}$ with a constant $v_f$ and setting $\eta_{vf} = 0$. Additionally, making the substitutions $Z_f \to 1$ and $g_\l \to g$ (which consequently sets $y = 0$) recovers the regularized fermionic bubble in the non-self-consistent calculation $\mathcal{X}_0(Q, \Lambda_f)$ used in section III. 

\noindent Our starting point is the full expression for the fermionic bubble diagram:
\begin{align}
    \mathcal{X}(Q, \lf) = 
    - \frac{2 g_\Lambda^2}{(2 \pi)^3} 
    \int_{-\infty}^{\infty} {\rm d}k_0 
    \int {\rm d}^2 k \;
      \partial_{\Lambda_f} R_f(\vec{k}) \; \big( G_f(K) \big)^2 
    \left[
        G_f(K+Q)+ G_f(K-Q)
    \right] \; ,
\end{align}
where the fermionic self-energy is expanded around Fermi surface as
\begin{align}
     G_f^{-1} = - i k_0 + \xi_{\vec{k}} + \Sigma(\vec{k}, k_0) \approx - i Z_{f} \, k_0 + \xi_{\vec{k}, \Lambda} \; .
\end{align}
Here $\xi_{\vec{k}, \Lambda}$ describes the spherical Fermi surface with flowing Fermi velocity $v_{f, \Lambda}$. Explicitly, the self-consistent bubble diagram is given by:
\begin{align}
    \mathcal{X}_{sc}(Q, \, \lf) = 
    - \frac{2 g_\Lambda^2}{(2 \pi)^3} 
    \int_{-\infty}^{\infty} {\rm d}k_0 
    \int {\rm d}^2 k 
    \frac{ \partial_{\Lambda_f} R_f(\vec{k}) }{ \big(-i Z_f k_0+ f(\vec{k}) \big)^2} 
    \Bigg[
        \frac{1}{-i Z_f (k_0 + q_0) + f(\vec{k} + \vec{q})}
        + 
        \frac{1}{-i Z_f (k_0 - q_0) + f(\vec{k} - \vec{q})}
    \Bigg] \; ,
\end{align}
where the regularized dispersion $f(\vec{k}) = \xi_{\vec{k}, \Lambda}+ R_f(\vec{k})$ incorporates the fermionic regulator.
While the frequency integral can be straightforwardly evaluated using the residue theorem, computing the momentum integrals is more involved. The phase space for momentum integration is tightly constrained by the scale derivative of the regulator, which reads:
\begin{align}\label{eq:DEF:pdRF}
    \partial_{\lf}R_f(\vec{k}) \approx 
    v_f \;
    \Bigg\{ &
     (1 - \eta_{vf}) \bigg( 
        \theta \big[ { p \; in \; shell} \big] - \theta \big[ { h \; in \; shell} \big] 
    \bigg) 
    + \eta_{vf} \frac{|\vec{k}|^2 - k_f^2}{2 k_f \Lambda_f} \bigg( 
        \theta \big[ { p \; in \; shell} \big] + \theta \big[ { h \; in \; shell} \big] 
    \bigg)
    \Bigg\} \; .
\end{align}
We neglect  contributions of order $\mathcal{O}(\Lambda_f/k_f)$ and we define the specific momentum-space regions\textit{"p in shell"} and \textit{"h in shell"} to denote particles and holes, respectively, located within a shell of width $\Lambda_f$ around the Fermi surface. Mathematically, these regions are defined via Heaviside step functions:
\begin{eqnarray}
    \theta \big[ { p \; in \; shell} \big] &\coloneqq& \theta \big[ \Lambda_f - (|\vec{k}| - k_f) \big] \; \theta \big[ |\vec{k}| - k_f  \big] \\ 
    \theta \big[ { h \; in \; shell} \big] &\coloneqq& \theta \big[ \Lambda_f - ( k_f - |\vec{k}|) \big]  \;\theta \big[  k_f - |\vec{k}| \big]  \; .
\end{eqnarray}
Changing variables $k_0 \to k_0/Z_f$ and evaluating the frequency integral via the residue theorem yields:
\begin{align}
    \mathcal{X}_{sc}(Q, \lf) = 
    &- \mathcal{A}_\Lambda \frac{v_{f, \Lambda}^2}{2 k_f}
    (1-\eta_{vf})
    \Bigg\{ 
    \int_{0}^{2 \pi}{\rm d}\phi
    \int_{k_f}^{k_f +\Lambda_f} {\rm d} k \, k
    \Bigg(
    \frac{\theta \big[ -f_{\Lambda}(\vec{k} + \vec{q}) \big] }{ \big(Z_f q_0 + i \xi_{k_f + \Lambda} - i f_{\Lambda}(\vec{k} + \vec{q}) \big)^2}  
    + (q_0 \to -  q_0) \Bigg) + \notag
     \\
    & \hspace{4. cm}+ 
    \int_{0}^{2 \pi}{\rm d}\phi
    \int_{k_f -\Lambda_f}^{k_f} {\rm d} k \, k \,
    \Bigg(
    \frac{  \theta \big[ f_{\Lambda}(\vec{k} + \vec{q}) \big]  }{ \big(Z_f q_0 + i f_{\Lambda}(\vec{k} + \vec{q}) - i \xi_{k_f - \Lambda} \big)^2}
    + (q_0 \to -  q_0)  \Bigg) 
    \Bigg\} + \notag
    \\
    &- \mathcal{A}_\Lambda \frac{v_{f, \Lambda}^2}{2 k_f} 
    \frac{\eta_{vf}}{2 k_f \Lambda_f} 
    \Bigg\{
    \int_{0}^{2 \pi}{\rm d}\phi
    \int_{k_f}^{k_f +\Lambda_f} {\rm d} k 
    \, k \,(k^2 - k_f^2) \,
    \Bigg(
    \frac{ \theta \big[ -f_{\Lambda}(\vec{k} + \vec{q}) \big]   }{ \big(Z_f q_0 + i \xi_{k_f + \Lambda} - i f_{\Lambda}(\vec{k} + \vec{q}) \big)^2} 
    + (q_0 \to -  q_0) \Bigg) +
    \notag \\ 
    & \hspace{3.7 cm}-
    \int_{0}^{2 \pi}{\rm d}\phi
    \int_{k_f -\Lambda_f}^{k_f} {\rm d} k
    \, k \,(k^2 - k_f^2) \,
    \Bigg(
    \frac{  \theta \big[ f_{\Lambda}(\vec{k} + \vec{q}) \big] }{ \big(Z_f q_0 + i f_{\Lambda}(\vec{k} + \vec{q}) - i \xi_{k_f - \Lambda} \big)^2} 
    + (q_0 \to -  q_0)
    \Bigg)
    \Bigg\} \; 
\end{align}
with a multiplication factor:
\begin{align} \label{constants}
    \mathcal{A}_\Lambda &= \frac{g_{\Lambda}^2 k_f}{ \pi^2  v_{f, \Lambda} Z_f} = 
    \mathcal{A}^* \,
    \Lambda_f^{- 2 \eta_g + \eta_{vf} + \eta_f} \; .
\end{align}

For convenience, we have not absorbed the factor $v_{f, \Lambda}^2/k_f$ (appearing in the first part of the expression) into the definition of $\mathcal{A}_\Lambda$. In the static limit ($q_0 = 0$), the denominators in the integrand are proportional to $v_{f, \Lambda} \Lambda_f$, since $f_{\Lambda}(\vec{k} + \vec{q}) - \xi_{k_f \pm \Lambda_f} \sim v_{f, \Lambda} \Lambda_f$; consequently, the Fermi velocity cancels out.
The additional factor of $k_f$ in the first part arises from the momentum integration. For momenta $\vec{k} +\vec{q}$ lying inside the $\Lambda_f$-shell around the Fermi surface, the integrals in the first two lines represent the areas of the \textit{particle-hole} and \textit{hole-particle} transfer regions (see, e.g., Fig.~\ref{fig:kregionsmall} for the specific case where $q < \Lambda_f$). A rough estimate of this area yields $\sim k_f \Lambda_f$.

\begin{figure}[h!] 
    \centering
    \includegraphics[width=0.4\linewidth]{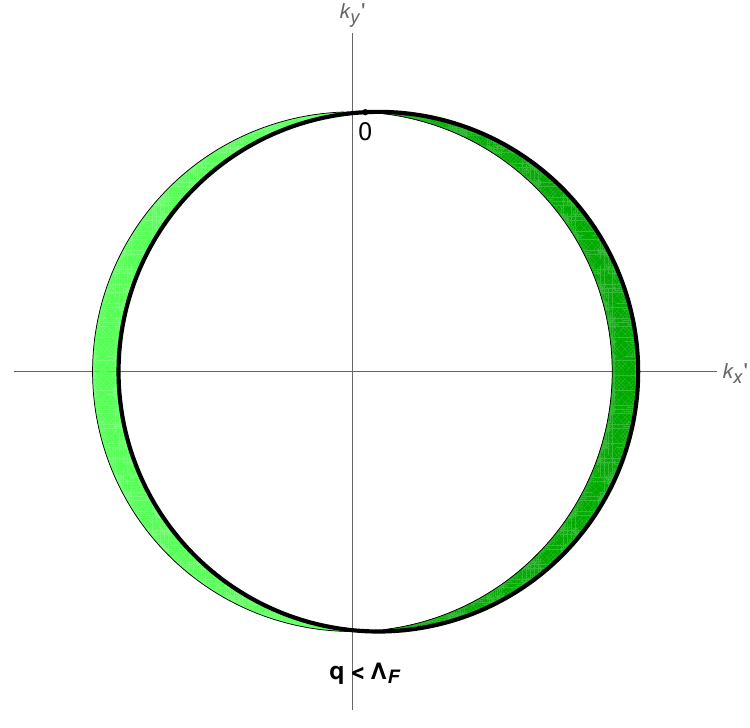}
    \caption{Regions of integration in momentum space for $|\vec{q}|<\Lambda_f$. Particles to holes transfers are shaded with lighter green color and holes to particles with darker green.}
    \label{fig:kregionsmall}
\end{figure}

\noindent The second part of the expression (proportional to $\eta_{vf}$) involves a factor of $v_f^2/k_f^2$. At first glance, this terms appears to be of higher order in the $\Lambda_f/k_f \ll 1$ limit because the third and fourth lines enter with opposite signs. However, estimating the magnitude of the momentum integrals for the \textit{particle-hole} and \textit{hole-particle} transfers inside the $\Lambda_f$-shell yields:
\begin{align}
    \int_{k_f}^{k_f + \Lambda_f} {\rm d}k \,(k^3 - k \, k_f^2) - \int_{k_f - \Lambda_f}^{k_f} {\rm d}k \,(k^3 - k \, k_f^2) 
    \approx 2 k_f^2 \Lambda_f^2 \; .
\end{align}
We therefore conclude that both parts --- the one proportional to $1 - \eta_{vf}$ and the one proportional to $\eta_{vf}$ --- are of the same  order in the limit $k_f \gg q \in \{\Lambda, \Lambda_f\}$.
By introducing dimensionless variables, the flowing bubble diagram can be rewritten as follows:
\begin{align}
    \mathcal{X}_{sc}(Q, \lf) &= 
    \frac{\mathcal{A}^*\Lambda_f^{y}}{q} \;
    \bigg[
     (1 - \eta_{vf}) \,
    h\big( t, w, \epsilon \big) 
    + \eta_{vf} \, h_{k^2}\big( t, w, \epsilon \big) 
    \bigg] \\[2ex]  
    &t \coloneqq \frac{\Lambda_f}{q} \;, \; w \coloneqq \frac{Z_f^* q_0}{ v_f^* q^x} \;, \; \epsilon \coloneqq \frac{q}{k_f}
\end{align}
Since we are interested in the limit of a large Fermi surface, we take $\epsilon \to 0$ to recover Eq.~(\ref{eq:SCFL:scalingCHI}). This allows us to define the scaling functions as:
\begin{equation}
    h(t, w) \coloneqq - \lim_{\epsilon \to 0}  \frac{q \, v_{f, \Lambda}^2}{2 k_f}
    \begin{multlined}[t]
    \Bigg\{ 
    \int_{0}^{2 \pi}{\rm d}\phi
    \int_{k_f}^{k_f +\Lambda_f} {\rm d} k \, k
    \Bigg(
    \frac{\theta \big[ -f_{\Lambda}(\vec{k} + \vec{q}) \big] }{ \big(Z_f q_0 + i \xi_{k_f + \Lambda} - i f_{\Lambda}(\vec{k} + \vec{q}) \big)^2}  
    + (q_0 \to -  q_0) \Bigg) + \\
    + 
    \int_{0}^{2 \pi}{\rm d}\phi
    \int_{k_f -\Lambda_f}^{k_f} {\rm d} k \, k \,
    \Bigg(
    \frac{  \theta \big[ f_{\Lambda}(\vec{k} + \vec{q}) \big]  }{ \big(Z_f q_0 + i f_{\Lambda}(\vec{k} + \vec{q}) - i \xi_{k_f - \Lambda} \big)^2}
    + (q_0 \to -  q_0)  \Bigg) 
    \Bigg\} \;, 
    \end{multlined}
\end{equation}
and 
\begin{equation}
    h_{k^2}(t, w) \coloneqq 
    - \lim_{\epsilon \to 0} \frac{q \, v_{f, \Lambda}^2}{4 k_f^2 \Lambda_f} 
    \begin{multlined}[t]
    \Bigg\{
    \int_{0}^{2 \pi}{\rm d}\phi
    \int_{k_f}^{k_f +\Lambda_f} {\rm d} k 
    \, k \,(k^2 - k_f^2) \,
    \Bigg(
    \frac{ \theta \big[ -f_{\Lambda}(\vec{k} + \vec{q}) \big]   }{ \big(Z_f q_0 + i \xi_{k_f + \Lambda} - i f_{\Lambda}(\vec{k} + \vec{q}) \big)^2} 
    + (q_0 \to -  q_0) \Bigg) +
    \\ 
    -
    \int_{0}^{2 \pi}{\rm d}\phi
    \int_{k_f -\Lambda_f}^{k_f} {\rm d} k
    \, k \,(k^2 - k_f^2) \,
    \Bigg(
    \frac{  \theta \big[ f_{\Lambda}(\vec{k} + \vec{q}) \big] }{ \big(Z_f q_0 + i f_{\Lambda}(\vec{k} + \vec{q}) - i \xi_{k_f - \Lambda} \big)^2} 
    + (q_0 \to -  q_0)
    \Bigg)
    \Bigg\} \; .
    \end{multlined}
\end{equation}
Finally, we point out that due to the non-analytic nature of the regulator $R_f$, the integrals must be evaluated separately for three distinct cases depending on the geometry shown in Fig.~\ref{fig:appb:kregions}. A detailed calculation for the non-self-consistent variant $\mathcal{X}_0$ can be found in the Supplemental Material of Ref.~\cite{Homenda_2024}.
\begin{align} \label{app:eq:regions}
\mathcal{X}_{sc}(Q, \lf) = 
    \begin{cases}
        \mathcal{X}_<(|\vec{q}|, q_0, \lf), \quad {\rm for} \; |\vec{q}| \leq \Lambda_f \\
        \mathcal{X}_M(|\vec{q}|, q_0, \lf), \quad {\rm for} \; |\vec{q}| \in (\Lambda_f, 2\Lambda_f) \\
        \mathcal{X}_>(|\vec{q}|, q_0, \lf), \quad {\rm for} \; |\vec{q}| \geq  2 \Lambda_f
    \end{cases} \; .
\end{align}

\begin{figure}[h]
    \centering
    \includegraphics[width=\textwidth]{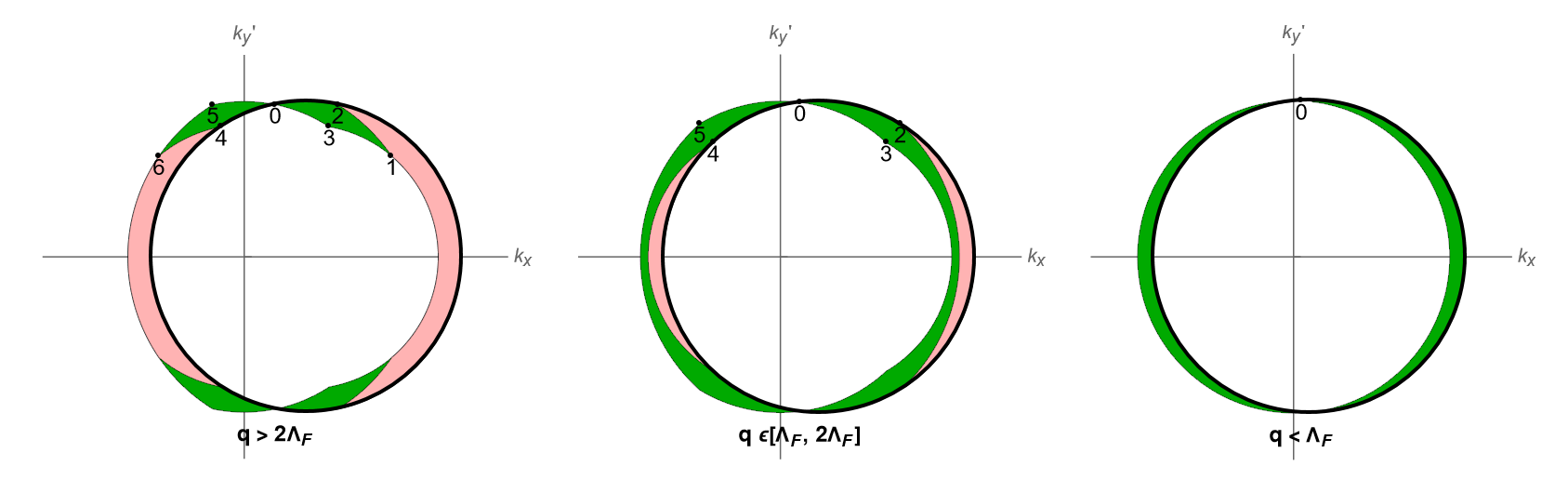}
\caption{ The regions of momentum space integration, which have non-zero contribution to the integrals. The integrand is constant for the momentum vector $\vec{k}$ which lies in the green region due to cut-off function which regularizes the fermi dispersion. For a momentum vector which lies in the red region, the cut-off function becomes inactive. Black points define the geometry of the region. On these plots we shifted the origin of the axis by $\vec{q}$, which is convenient for calculations. } 
\label{fig:appb:kregions}
\end{figure}

\subsection{$h(t, w)$ and $h_{k^2}(t, w)$}
We present the scaling function after evaluating the momentum integration according to the notation in Eq.~(\ref{app:eq:regions}):
\begin{align}
h_{<}(t, w) ={}& 
\dfrac{1}{t^2} \dfrac{16 - 4 w^2 t^{-2x}}{(4 + w^2 t^{-2x})^2}  \\
h_{M}(t, w) ={}&
\begin{aligned}[t]
\dfrac{1}{t^2} \dfrac{16 - 4 w^2 t^{-2x}}{(4 + w^2 t^{-2x})^2} \Big( 1 - 2 \sqrt{1 - t^2} + 2 t \arccos(t) \Big) 
- \frac{4}{t}\int_0^{\arccos(t)} {\rm d}\phi \Big(
\dfrac{1 + t^{-1}   \cos(\phi)}{ w^2 t^{-2x} + (1 + t^{-1}   \cos(\phi))^2} + \\
 - \dfrac{2}{4 + w^2 t^{-2x} }
\Big)
\end{aligned}
\\
h_{>}(t, w) ={}&
\begin{aligned}[t]
\dfrac{1}{t^2} \dfrac{16 - 4 w^2 t^{-2x}}{(4 + w^2 t^{-2x})^2} 
\Big(
1 + \sqrt{1 - 4 t^2} - 2 \sqrt{1 - t^2} - 2 t (\arccos(2t) - \arccos(t))
\Big) 
+\frac{8}{t} \frac{\arccos(t) - \arccos(2t)}{4 + w^2 t^{-2x}} + \\ + \frac{4}{\sqrt{1 + w^2 t^{2-2x}}} \arctan \Bigg(\sqrt{\frac{1 - 4 t^2}{1+ w^2 t^{2-2x}}} \Bigg)
- \dfrac{4}{t^2}\int_0^{\arccos(t)} {\rm d}\phi 
\dfrac{t + \cos(\phi)}{ w^2 t^{-2x} + (1 + t^{-1} \cos(\phi))^2} 
\end{aligned}
\end{align}
and
\begin{align}
h_{k^2,<}(t, w) &=
\dfrac{\pi}{2 t^3} \dfrac{4 - w^2 t^{-2x}}{\big(4 + w^2 t^{-2x} \big)^2}
 \\
h_{k^2,M}(t, w) &= 
\!\begin{aligned}[t]
\dfrac{\pi + 4 t \sqrt{1 - t^2} - 4 \arccos(t)}{2t^3} 
\, \dfrac{4 - w^2 t^{-2x}}{\big(4 + w^2 t^{-2x} \big)^2}  
- \dfrac{2}{t}\int_{0}^{\arccos(t)}{\rm d}\phi \ln \left( \dfrac{w^2 t^{2 - 2x} + (t + \cos(\phi))^2}{w^2 t^{2 - 2x}+ 4 t^2} \right) + \\
+ \dfrac{8}{t^2}
\dfrac{\sqrt{1 - t^2} -t \arccos(t)}{4 + w^2 t^{-2x}} \\
\end{aligned}
\\
h_{k^2,>}(t, w) &=
\!\begin{aligned}[t] \label{app:hB}
\dfrac{1}{t^3} \dfrac{4 -  w^2 t^{-2x}}{\big(4 + w^2 t^{-2x} \big)^2}
\Big(\dfrac{2t (1 - t^2  - \sqrt{1 - 5 t^2 + 4 t^4})}{\sqrt{1 - t^2}} + 2\arcsin(t) - \arcsin(2t) \Big) 
\; +\\
  - \frac{2}{t} \int_0^{\arccos(2t)} {\rm d} \phi \left[
\ln \left( \dfrac{w^2 t^{2 - 2 x} + \big(t + \cos(\phi) \big)^2}{ w^2 t^{2 - 2x} +\cos^2(\phi)} \right)
- \dfrac{2 t \cos(\phi)}{w^2 t^{2 - 2x} + \cos^2(\phi)}
\right]
\; + \\ 
  -\dfrac{2}{t} \int_{\arccos(2t)}^{\arccos(t)}{\rm d}\phi
\left[ 
\ln\left( \dfrac{w^2 t^{-2x} + \big(1 + t^{-1}cos(\phi) \big)^2}{4 + w^2 t^{-2x}} \right)
+
\dfrac{4(t - \cos(\phi))}{t(4 + w^2 t^{-2x})}
\right] \; .
\end{aligned}
\end{align}

\subsection{Static limit: $h_{k^2}(t)$ and $h_{k^2}(t)$} \label{sec:appendix:scaling}

Below we present static limit of the scaling functions, $h(t)\coloneq h(t, w = 0)$ and $h_{k^2}\coloneq h_{k^2}(t, w = 0)$: 
\begin{align} \label{eq:S3:h(t)}
    h(t) = 
    \begin{cases}
    \dfrac{1}{t^2} \;, &{\rm for} \; t \geq 1 \\[1.2ex]
    \dfrac{1}{t^2} \; (1 - 2 \sqrt{1 - t^2}) + \dfrac{4}{t} \; \arccos(t) + 4\dfrac{\ln(t)}{\sqrt{1 - t^2}} 
    \; , &{\rm for} \; t \in (\frac{1}{2}, 1) \\[1.2ex]
    4 {\rm arctanh}(\sqrt{1 - t^2}) + 4\dfrac{\ln(t)}{\sqrt{1 - t^2}} + 4 \big(\arccos(t) - \arccos(2t)\big) - \dfrac{1}{t^2} \big(2\sqrt{1- t^2} - \sqrt{1- 4t^2} - 1 \big) \; , & {\rm for} \; t \leq \frac{1}{2}
    \end{cases} \; ,
\end{align}
and
\begin{align}
    h_{k^2}(t) = 
    \begin{cases}
    \dfrac{\pi}{8 t^3} \;, &{\rm for} \; t \geq 1 \\[1.2 ex]
    \dfrac{1}{8 t^3} \left( \pi + 4 t \sqrt{1 - t^2} - 4 \arccos(t) \right)        + \dfrac{2}{t^2} \left( \sqrt{1 - t^2} - t \arccos(t)  \right)  - \dfrac{4}{t} \int_{0}^{\arccos(t)} {\rm d} \phi \ln\big( \dfrac{t + \cos(\phi)}{2t} \big)
    \, , \quad &{\rm for} \; t \in (\frac{1}{2}, 1)\\[1.2 ex]
    \dfrac{5 (1 - t^2 - \sqrt{1 - 5 t^2 + 4 t^4})}{2t^2 \sqrt{1 - t^2}}
    + \big(\arcsin(t) - \arcsin(2t) \big) \dfrac{1 + 8 t^2 - 16 t^2 \ln(2t) }{4 t^3}
    + \dfrac{\arcsin(t)}{4 t^3} + \\
    \hfill+ 4 \,{\rm arctanh}(\sqrt{1 - 4 t^2}) 
    -\dfrac{4}{t} \int_0^{\arccos(t)}{\rm d}\phi \ln \big(t + \cos (\phi) \big) + 
    \dfrac{4}{t} \int_0^{\arccos(2t)}{\rm d}\phi \ln \big( \cos (\phi) \big)
    \; , & {\rm for} \; t \leq \frac{1}{2}
    \end{cases} \; .
\end{align}
Both functions have finite $t\to0$ ($q\gg \lf$) limit:
\begin{align} \label{app:staticExp}
    &h(t\approx 0) = 3 + \mathcal{O}(t^2) \\
    &h_{k^2}(t\approx 0) = \frac{3}{2} + \mathcal{O}(t^2) \; .
\end{align}

\section{$B_0(q_0, q, \Lambda_f)$ for small $\epsilon_\l$.}

We start from the Eq.~(\ref{B000}). Since the function $h(t,w)$ is an analytical function for any $t>0$, then $B_0(q_0, q, \lf)$ is also an analytical function of $q_0$ unless we take $\epsilon_\l\to0$ ($Q_\l\gg \lf$) in the integration limit. Therefore we study the behavior of $h_>(t, w)$ for $t\to0$. We recognize that the relevant term as the first line in Eq.~(\ref{app:hB}), which is factorized into the static and dynamical part. For $t\ll1$ the static part can be expanded in $t$ and the subsequent terms take the form:
\begin{align}
&\dfrac{1}{t^2} \dfrac{16 - 4 w^2 t^{-2}}{(4 + w^2 t^{-2})^2} 
\Big(
1 + \sqrt{1 - 4 t^2} - 2 \sqrt{1 - t^2} - 2 t (\arccos(2t) - \arccos(t))
\Big) \approx \dfrac{16 - 4 w^2 t^{-2}}{(4 + w^2 t^{-2})^2} (1 + \mathcal{O}(t^2)) \\
&\frac{8}{t} \frac{\arccos(t) - \arccos(2t)}{4 + w^2 t^{-2x}} \approx \frac{8}{4 + w^2 t^{-2x}}(1 + \mathcal{O}(t^2)) \; .
\end{align}
We approximate the scaling function by:
\begin{align}
    h_>(t, w)\approx \frac{16 - 4 w^2 t^{-2}}{(4 + w^2 t^{-2})^2} +
     \frac{8}{4 + w^2 t^{-2x}} \; ,
\end{align}
which posses a discontinuity at $\{ t = 0, w = 0 \}$:
\begin{align}
\!\begin{aligned}[t]
    \lim_{t\to0} \lim_{w\to0} h(t, w) = 3 \\
    \lim_{w\to0} \lim_{t\to0} h(t, w) = 0 \; .
\end{aligned}
\end{align}
The higher order terms in $t$ do not play any role. Therefore we plug the expanded $h_>(t,w)$ function into Eq.~(\ref{B000}) and after integration over $t$ we obtain the expression in Eq.~(\ref{eq:NSC:Bdynamic}):
\begin{align}
    \int_{1/2}^{\frac{\lf}{Q_\l}}{\rm d} t \,
    \left(
    \frac{16 - 4 w^2 t^{-2}}{(4 + w^2 t^{-2})^2} +
     \frac{8}{4 + w^2 t^{-2x}}
    \right) = 
    \Bigg[ 3 t + \frac{t \, w^2}{4 t^2 + w^2} - 2 w \arctan \Big( \frac{2t}{w}\Big) \Bigg]_{t = 1/2}^{t = \lf/Q_\l} \; .
\end{align}
Using Eq.~(\ref{app:staticExp}) we can easily retrieve Eq.~(\ref{rob}):
\begin{eqnarray}
    B_{0}(0,Q_\Lambda, \Lambda) = \mathcal{A}\int_{\infty}^{\Lambda_f} {\rm d} \lambda \;
     \frac{h(\lambda/Q_{\Lambda})}{Q_{\Lambda}} =
     \mathcal{A} \left(
     \int_{\infty}^0{\rm d}t  \;h(t) + 
     \int_{0}^{\frac{\Lambda_f}{Q_\Lambda}}{\rm d}t  \;h(t)
     \right)
     \approx \Delta m_{b,0}^2 + 3 \frac{\mathcal{A} \; \Lambda_f}{Q_\Lambda} = \Delta m_{b,0}^2 + 2 Q_\Lambda^2.
\end{eqnarray}
To recover the Hertz-Millis theory [see Eq.~(\ref{B_w})], we use a finite $q$ instead of $Q_\l$, keeping in mind that $\lf \to 0$ much faster than $q$. This validates the expansion of the static factors in $h_>(t,w)$ with respect to $t$. By construction, the Hertz-Millis framework encounters the problem of a vanishing flowing mass term, since $\lf = 0$ while $\l \neq 0$.

\section{Flowing Fock diagram }

We calculate the momentum integrals occurring in Eq.~(\ref{eq:freq:sigmaflow}). 
We focus on the self-energy at  the Fermi surface, $\Sigma(k_0) = \Sigma(k_0, \vec{k} = k_f \hat{e}_x)$: 
\begin{eqnarray}
    \partial_{\lf} \Sigma(k_0) =
    \frac{g^2}{8 \pi^3} 
    \Bigg( 
        \int_Q \partial_{\lf}R_b(q)  \; G_b^2(Q) G_f(K_Q)
        +\int_Q G_b(Q) \partial_{\lf} R_f(k_f \hat{e}_x + \vec{q}) \; G_f^2(K_Q)
    \Bigg) \,,
\end{eqnarray}
where $K_Q = \{ k_0 -q_0, k_f \hat{e}_x + \vec{q}\}$.
After applying the fermionic cut-off function introduced in Eq.~(\ref{eq:Rf}) and bosonic regulator $R_b = (\l^2 - (q - Q_\l)^2) \theta[\l^2 - (q - Q_\l)^2]$ we find:

\begin{eqnarray}
    \partial_{\lf} \Sigma(k_0) =
    \frac{g^2}{8 \pi^3}
    \int_{-\infty}^{\infty} {\rm d}q_0
    \Bigg(
        \int_0^{\l + Q_\l} {\rm d}q \;
        \frac{2 q (\partial_\lf \l) \big( \l + \partial_\l Q_\l (q - Q_\l) \big)}{\left( m_{b}^2 + \l^2 + (q_0 + k_0)^2 + B(q_0 + k_0, q, \lf) \right)^2}
        \int_0^{2 \pi} {\rm d} \phi
        \frac{1}{ i q_0 + f(k_f + \vec{q}) + \Sigma(- q_0) } \nonumber \\
        + 
        \bigg(\int_{q_{out}} 
        \frac{1}{m_{b}^2 + (|\vec{q}| - Q_\l)^2 + (q_0 + k_0)^2 + B(q_0 + k_0, |\vec{q}|, \lf) + R_b(|\vec{q}|) }
        \bigg)
        \frac{v_f\left(1 + \frac{\lf}{k_f}\right)}{\left( i q_0 + v_f \lf (1 + \frac{\lf}{k_f}) + \Sigma(- q_0) \right)^2} \nonumber \\
        - 
        \bigg(\int_{q_{in}} 
        \frac{1}{m_{b}^2 + (|\vec{q}| - Q_\l)^2 + (q_0 + k_0)^2 + B(q_0 + k_0, |\vec{q}|, \lf) + R_b(|\vec{q}|) }
        \bigg)
        \frac{v_f\left(1 - \frac{\lf}{k_f}\right)}{\left( i q_0 + v_f \lf (1 - \frac{\lf}{k_f}) + \Sigma(- q_0) \right)^2}
    \Bigg) \; ,\nonumber \\
\end{eqnarray}
where $f(\vec{k}) = \xi_{\vec{k}} + R_f(\vec{k})$ denotes the regularized fermionic dispersion. We also shifted the frequency domain by $- k_0$. The integral subscripts in the second and third lines denote the integration regions in the  $\vec{q}$ space, which are shown in  Fig.~\ref{fig:qoutin}. The vector $\vec{q}_{in}$ denotes the momentum for which fermionic momentum $k_f \hat{e}_x + \vec{q}$ lies inside the Fermi sea, while for $\vec{q}_{out}$ fermionic momentum exceeds $k_f$.
We  neglect the real part of the fermionic self-energy and rewrite the flow equation for the imaginary part using $\tilde{\Sigma} (k_0) \coloneqq k_0 - \Im (\Sigma(k_0))$:
\begin{eqnarray}
    \partial_{\lf} \Im(\Sigma(k_0)) =
    - \frac{g^2}{8 \pi^3}
    \int_{-\infty}^{\infty} {\rm d}q_0
    \Bigg(
        \int_0^{\l + Q_\l} {\rm d}q \;
        \frac{2 q (\partial_\lf \l) \big( \l + \partial_\l Q_\l (q - Q_\l) \big)}{\left( m_{b}^2 + \l^2 + (q_0 + k_0)^2 + B(q_0 + k_0, q, \lf) \right)^2}
        \int_0^{2 \pi} {\rm d} \phi
        \frac{\tilde{\Sigma}(q_0)}{ \tilde{\Sigma}(q_0)^2 + f(k_f + \vec{q})^2 } \nonumber\\
        + 
        \bigg(\int_{q_{out}} 
        \frac{1}{m_{b}^2 + (|\vec{q}| - Q_\l)^2 + (q_0 + k_0)^2 + B(q_0 + k_0, |\vec{q}|, \lf) + R_b(|\vec{q}|) }
        \bigg)
        \frac{v_f\left(1 + \frac{\lf}{k_f}\right) \tilde{\Sigma}(q_0)}{ \left( \tilde{\Sigma}(q_0)^2 + v_f^2 \lf^2 (1 + \frac{\lf}{k_f})^2 \right)^2 } \nonumber\\
        - 
        \bigg(\int_{q_{in}} 
        \frac{1}{m_{b}^2 + (|\vec{q}| - Q_\l)^2 + (q_0 + k_0)^2 + B(q_0 + k_0, |\vec{q}|, \lf) + R_b(|\vec{q}|) }
        \bigg)
        \frac{v_f\left(1 - \frac{\lf}{k_f}\right) \tilde{\Sigma}(q_0)}{ \left( \tilde{\Sigma}(q_0)^2 + v_f^2 \lf^2 (1 - \frac{\lf}{k_f})^2 \right)^2 }
    \Bigg) \; . \nonumber\\
\end{eqnarray}
We used the fact that $\Im (\Sigma(-q_0)) = -\Im(\Sigma(q_0))$. 
After the angular integration and expansion of the $B(Q,\lf)$:
\begin{align} \label{eq:detailSigma}
    \partial_\lf \Im(\Sigma(k_0)) &=
    - \frac{g^2}{2 \pi^3} \int_0^{\infty}{\rm d}q_0 \;
    \Bigg\{
        \int_0^{\lf} {\rm d} q\;
        \frac{\pi q (\partial_\lf \l) \big(\l + \partial_\l Q_\l (q - Q_\l)\big)}{\Big(m_b^2 + 3 Q_\l^2 + q_0^2 + \mathcal{A} \pi \frac{|q_0|}{v_f Q_\l} \Big)^2}
        \Big(L_1(q_0 - k_0, v_f \lf) - L_1(q_0 + k_0, v_f \lf) \Big) + \notag \\ \notag
        & +
        \int_\lf^{\l + Q_\l} {\rm d} q\;
        \frac{ q (\phi_3 - \phi_1) (\partial_\lf \l) \big(\l + \partial_\l Q_\l (q - Q_\l)\big)}{\Big(m_b^2 + 3 Q_\l^2 + q_0^2 + \mathcal{A} \pi \frac{|q_0|}{v_f Q_\l} \Big)^2}
        \Big(L_1(q_0 - k_0, v_f \lf) - L_1(q_0 + k_0, v_f \lf) \Big) + \\
        & + 2
        \int_\lf^{\l + Q_\l} {\rm d} q\;
        \frac{ (\partial_\lf \l) \big(\l + \partial_\l Q_\l (q - Q_\l)\big)}{v_f \Big(m_b^2 + 3 Q_\l^2 + q_0^2 + \mathcal{A} \pi \frac{|q_0|}{v_f Q_\l} \Big)^2}
        \Bigg(I_0 \Big(\frac{q_0 - k_0}{v_f q}, \phi_1 \Big) - I_0 \Big(\frac{q_0 + k_0}{v_f q}, \phi_1 \Big) \Bigg) +  \\ 
        & + 
        \int_0^{\lf} {\rm d} q\;
        \frac{\pi v_f^2 q \lf }{m_b^2 + 3 Q_\l^2 + q_0^2 + \mathcal{A} \pi \frac{|q_0|}{v_f Q_\l} }
        \Big(L_2(q_0 - k_0, v_f \lf) - L_2(q_0 + k_0, v_f \lf) \Big) + \notag \\ \notag
        & + 
        \int_{\lf}^{\l + Q_l} {\rm d} q\;
        \frac{ (\phi_3 - \phi_1) v_f^2 q \lf }{m_b^2 + 3 Q_\l^2 + q_0^2 + \mathcal{A} \pi \frac{|q_0|}{v_f Q_\l} }
        \Big(L_2(q_0 - k_0, v_f \lf) - L_2(q_0 + k_0, v_f \lf) \Big)
    \Bigg\} \; .
\end{align}
We focus on the large  $k_f$ limit, where $\pi - \phi_3 \to \phi_1$. The definitions of the function $L_n$ and $I_0$ are presented below:
\begin{subequations} \label{eq:definitions}
\begin{eqnarray}
    && \phi_1 \coloneqq \arccos \left( \frac{\lf}{q} + \frac{\lf^2}{2 k_f q} - \frac{q}{2 k_f} \right) 
 \xrightarrow{k_f \gg \l, \lf} \arccos \left( \frac{\lf}{q}\right) \\
     && \phi_3 \coloneqq \arccos \left( - \frac{\lf}{q} + \frac{\lf^2}{2 k_f q} - \frac{q}{2 k_f} \right) 
 \xrightarrow{k_f \gg \l, \lf} \arccos \left( - \frac{\lf}{q}\right) \\
    &&L_n(a, b) \coloneqq \frac{a}{(a^2 + b^2)^n} \\
    &&I_0(a, \phi) \coloneqq \int_0^{\phi} {\rm d} \phi' \frac{a}{a^2 + \cos^2(\phi')} = \frac{\arctan \left( \frac{a \tan(\phi)}{\sqrt{1 + a^2}} \right)}{\sqrt{1 + a^2}} \; .
\end{eqnarray}
\end{subequations}

\begin{figure}[h!]
\includegraphics[width =  0.3 \linewidth]{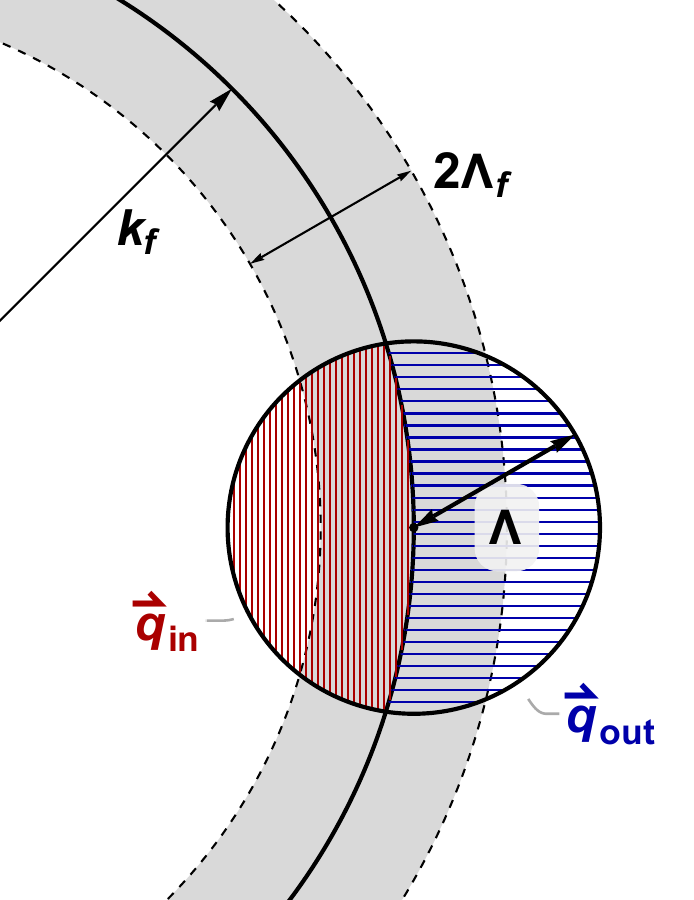}
\caption{ Part of the regularized Fermi surface with marked regions of internal integration yielding nonvanisahing contributions. Here $\vec{q}_{in}$ denotes all vectors for which $|k_f \hat{e}_x + \vec{q}| < k_f$ and $\vec{q}_{out}$ for which $|k_f \hat{e}_x + \vec{q}| > k_f$. } 
\label{fig:qoutin}
\end{figure}

\section{Numerical details}
Numerical results of the flow of fermionic self-energy presented in Sec.~III.B. involves two components:  solving a set of nonlinear differential equations and evaluation of 2-dimensional internal integrals changing with the flow.

To tackle the first task we used the embedded Runge-Kutta method with adaptive step size to minimize the number of RHS evaluations. 
%Moreover one can expect rather small changes of the parameters at the beginning of the flow, since the bosonic mass is nonzero. 
The absolute and relative tolerances which dictate the variation of step size were set to $10^{-12}$ and $10^{-13}$ respectively.
Disregarding the fermionic self-energy in the flow equation for bosonic correlation function allows to calculate the flow of bosonic quantities, tune to the criticality with desired precision and then calculate the flow of the fermionic self-energy taking into account previously obtained flow of the bosonic parameters. We implemented an adaptive step size algorithm for the fermionic sector in  a way that allows, if necessary, to use additional steps between the time points implemented for the bosonic flow. Therefore the fermionic step size was always smaller or equal to the bosonic step size.

Evaluating the integrals  required a detailed preceding analysis, especially concerning the $\l$-dependence of the integrands. In Eq.~(\ref{eq:detailSigma}) the integrands viewed as functions  of internal frequency $q_0$ exhibit peaks located at external frequency $k_0$ of the width of order $\l^2$ or $\l^3$ depending on the external frequency. The frequency integral can be schematically written as
\begin{eqnarray}
    I_{q_0} = \int_0^{\infty}{\rm d}q_0 
    \left(
    f(q_0 + k_0) - f(q_0 - k_0)
    \right)
    g(q_0) \;,
\end{eqnarray}
where $f(q_0)$ is even with respect to its internal argument while $g(q_0)$ is odd. Now we change the variables in such a way that the peak of the integrand is  located at the origin:
\begin{eqnarray} \label{eq:changeofvariables}
    I &=& 
    \int_0^{\infty}{\rm d}\tilde{q}_0 \, f\big(\tilde{q}_0 + k_0 \big) 
    \left(
    g\big(\tilde{q}_0\big) - g\big(\tilde{q}_0 + 2 k_0\big)
    \right)
     - k_0 \int_0^{1}{\rm d} t_{q0} \, f \big( t_{q0} k_0 \big) 
    g \big( k_0 (t_{q0}+1) \big) \\
    && - \frac{k_0}{2} \int_0^{1}{\rm d} t_{q0} \, f\big( \tfrac{k_0}{2} t_{q0} \big) 
    g\big( \tfrac{k_0}{2} (2- t_{q0}) \big)
    -\frac{k_0}{2} \int_0^{1}{\rm d} t_{q0} \, f\big( \tfrac{k_0}{2} (2 - t_{q0}) \big) 
    g\big( \tfrac{k_0}{2} t_{q0} \big) \; .
\end{eqnarray}
We split the integration domain into high frequency region with $\tilde{q}_0>5$ and low frequency region $\tilde{q}_0 <5$. The high frequency tail we evaluate after performing a substitution:
\begin{eqnarray}
    x = \tan \left(
    0.5(\frac{\pi}{2} - \arctan(5) )(\tilde{q}_0 + 1) \right) \; .
\end{eqnarray}
The low frequency region we divide into smaller subregions on logarithic scale. For each of the regions we implement Gauss-Legendre quadrature with 60-200 nodes.

\section{Tadpole auxiliary function}

The function $f(t)$ we define as [see Eq.~(\ref{eq:xigamma})]:
\begin{equation} \label{eq:ROUTE:fdefinition}
    f \Big( \frac{\lambda}{q} \Big) \coloneqq - i \frac{Z_{f,\lambda}^2 \pi^2 q}{v_f} 
    \Bigg\{
    2 \int_K  k_0 \,
     S_{f, \lambda}(K) G_{f}(K, \lambda)
    \Big[ G_{f}(K + Q, \lambda) + G_{f}(K - Q, \lambda)
    \Big]
    + \int_K  k_0 \,
    G_{f}^2 (K, \lambda)
    \Big[ S_{f}(K + Q, \lambda) + S_{f}(K - Q, \lambda)
    \Big]
    \Bigg\}
\end{equation}

We split the $f(t)$ function into two parts:
\begin{align}
    f_1(t) &\coloneq
    - i \frac{Z_{f,\lambda}^2 \pi^2 q}{v_f} 
    2 \int_K  k_0 \,
     S_{f}(K, \lambda) G_{f}(K, \lambda)
    \Big[ G_{f}(K + Q, \lambda) + G_{f}(K - Q, \lambda)
    \Big] \\
    f_2(t) &\coloneq
    - i \frac{Z_{f, \lambda}^2 \pi^2 q}{v_f} 
    \int_K  k_0 \,
    G_{f}^2 (K, \lambda)
    \Big[ S_{f}(K + Q, \lambda) + S_{f}(K - Q, \lambda)
    \Big]\;.
\end{align}

The final form is shown as:

\begin{align}
    f_1(t) = 
    \begin{cases}
    \dfrac{1}{4 t^2} \;, &{\rm for} \; t \geq 1 \\[1.2 ex]
    \dfrac{1}{4 t^2} \Big( 1 - 2 \sqrt{1 - t^2} + 2 t \arccos(t) \Big) 
    + \dfrac{3 \arccos(t)}{4 t} - \dfrac{2 \big(-1 + t^2 + 2 (-2 + t^2) \log(t) \big)}{4 (1 - t^2)^{3/2}}
    \, , \quad &{\rm for} \; t \in (\frac{1}{2}, 1)\\[1.2 ex]
    -\dfrac{1}{4} \Bigg( 2 \sqrt{1 - 4 t^2} - \dfrac{2}{\sqrt{1 - t^2}} 
    + \dfrac{-1 - \sqrt{1 - 4 t^2} + 2 \sqrt{1 - t^2} + 2 t \arccos(2t) - 2 t \arccos(t)}{t^2} \Bigg) + \\
    \hfill - \dfrac{3 \big( \arcsin(t) - \arcsin(2t) \big)}{4t} 
    + 4 \, {\rm arctanh}\bigg( \sqrt{-1 + \dfrac{2}{1 + 2t}} \bigg) 
    + \dfrac{ (2 - t^2) \log(t)}{(1 - t^2)^{3/2}}
    \; , & {\rm for} \; t \leq \frac{1}{2} 
    \end{cases} 
\end{align}
and
\begin{align}
    f_2(t) = 
    \begin{cases}
    0 \;, &{\rm for} \; t \geq 1 \\[1.2 ex]
    \dfrac{\arccos(t)}{4 t} + \dfrac{1 - t^2 + 2 \ln(t)}{2 (1 - t^2)^{3/2}}
    \, , \quad &{\rm for} \; t \in (\frac{1}{2}, 1)\\[1.2 ex]
    \dfrac{1}{2} \Bigg( \dfrac{\arcsin(2t) - \arcsin(t)}{2t} + \dfrac{1}{\sqrt{1 - t^2}} - \sqrt{1 - 4 t^2}  + \dfrac{2 \ln(t)}{(1 - t^2)^{3/2}} + 4 \, {\rm arctanh}\bigg( \sqrt{\dfrac{1 - 2 t}{1 + 2 t}} \bigg) \Bigg)
    \; , & {\rm for} \; t \leq \frac{1}{2} 
    \end{cases} \; .
\end{align}

\begin{figure}[h!] \label{fig:S4:ffunctions}
    \centering
    \includegraphics[width=0.5\linewidth]{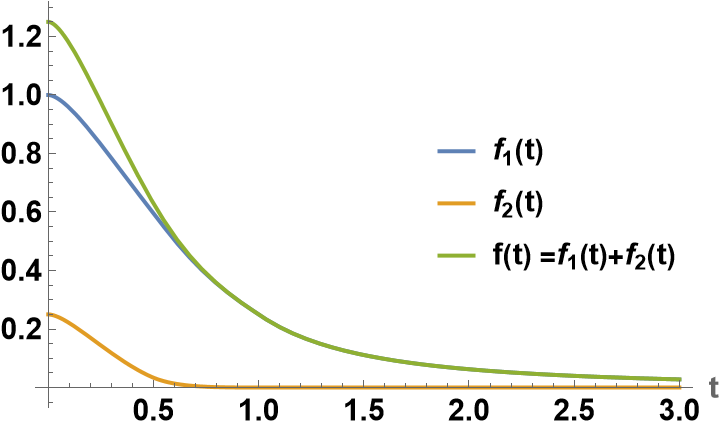}
    \caption{Plot of the scaling function important for the analysis in Sec.~IV.C. }
\end{figure}

\end{widetext}

\end{document}